\documentclass{article}

\usepackage{arxiv}
\usepackage[utf8]{inputenc} % allow utf-8 input

\usepackage{amsmath}
\usepackage{bm}
\usepackage{enumitem}
\usepackage{float}
\usepackage{natbib}
\usepackage{caption}
\usepackage{subcaption}
\usepackage{array}
\usepackage{amsfonts}
\usepackage[dvipsnames]{xcolor}
\usepackage{booktabs}
\usepackage{nomencl}
\makenomenclature
\usepackage{soul}
\usepackage{arydshln}
\usepackage{graphicx}
\usepackage{placeins}
\graphicspath{ {./images/} }

\usepackage{algorithm}
\usepackage[noend]{algpseudocode}

\setcitestyle{numbers} %Cite as numbers or author-year.
\usepackage{tikz}
\usepackage{pgfplots}
\DeclareRobustCommand\full{\tikz[baseline=-0.6ex]\draw[thick] (0,0)--(0.5,0);}
\DeclareRobustCommand\dotted{\tikz[baseline=-0.6ex]\draw[thick,dotted] (0,0)--(0.54,0);}
\DeclareRobustCommand\dashed{\tikz[baseline=-0.6ex]\draw[thick,dashed] (0,0)--(0.54,0);}
\DeclareRobustCommand\dotdash{\tikz[baseline=-0.6ex]\draw[thick,dash dot] (0,0)--(0.5,0);}
\usetikzlibrary{shapes.misc}
\tikzset{cross/.style={cross out, draw=black, minimum size=2*(#1-\pgflinewidth), inner sep=0pt, outer sep=0pt},
cross/.default={2.5pt}}
\DeclareRobustCommand\markercircle{\raisebox{0.5pt}{\tikz{\node[draw,scale=0.4,circle,fill=black](){};}}}
\DeclareRobustCommand\markercross{\raisebox{0.25pt}{\tikz{\node[draw,cross](){};}}}

\title{Probabilistic Machine Learning to Improve Generalisation of Data-Driven Turbulence Modelling}
\author{
    Joel Ho \\
    Independent Researcher \\
    Oxford, UK \\
    \texttt{joel.ho@hotmail.com}
    \And 
    Nick Pepper \\
	The Alan Turing Institute\\
	The British Library\\
	London, UK \\
	\texttt{npepper@turing.ac.uk} \\
    \And
    Tim Dodwell  \\
	Department of Computer Science\\
	University of Exeter\\
	Exeter, UK \\
}
\begin{document}

\maketitle

% Uncomment to remove the date
%\date{}

% Uncomment to override  the `A preprint' in the header
%\renewcommand{\headeright}{Technical Report}
%\renewcommand{\undertitle}{Technical Report}
\renewcommand{\shorttitle}{Prob. ML to Improve Generalisation of Data-Driven Turbulence Modelling}

%%% Add PDF metadata to help others organize their library
%%% Once the PDF is generated, you can check the metadata with
%%% $ pdfinfo template.pdf

%%%% Article title to be placed here

%\author{Nick Pepper$^{1}$, Marc Thomas$^{2}$, George {De~Ath}$^{3}$, Enrico Oliver$^{3}$, Richard Cannon$^{2}$, Richard Everson$^{3}$ and Tim Dodwell$^{3,4}$}

%% or include affiliations in footnotes:
%\address{$^{1}$ The Alan Turing Institute, The British Library, London, UK. \\
%$^{2}$ NATS \\
%$^{3}$ Department of Computer Science, University of Exeter, Exeter, UK. \\
%$^{4}$ digiLab, Exeter, UK.}

%%%% Keyword entries to be placed here %%%%
%\keywords{Trajectory Prediction, Probabilistic Machine Learning, Functional Data Analysis, Gaussian Process Emulators, Monotonicity}

%%%% Insert corresponding author and its email address}
%\corres{Nick Pepper\\
%\email{npepper@turing.ac.uk}}

%%%% Abstract text to be placed here %%%%%%%%%%%%
\begin{abstract}
A probabilistic machine learning model is introduced to augment the $k-\omega\ SST$ turbulence model in order to improve the modelling of separated flows and the generalisability of learnt corrections. Increasingly, machine learning methods have been used to leverage experimental and high-fidelity data, improving the accuracy of the Reynolds Averaged Navier Stokes (RANS) turbulence models widely used in industry. A significant challenge for such methods is their ability to generalise to unseen geometries and flow conditions. Furthermore, heterogeneous datasets containing a mix of experimental and simulation data must be efficiently handled. In this work, field inversion and an ensemble of Gaussian Process Emulators (GPEs) is employed to address both of these challenges. The ensemble model is applied to a range of benchmark test cases, demonstrating improved turbulence modelling for cases with separated flows with adverse pressure gradients, where RANS simulations are understood to be unreliable. Perhaps more significantly, the simulation reverted to the uncorrected model in regions of the flow exhibiting physics outside of the training data. 

\end{abstract}
%%%%%%%%%%%%%%%%%%%%%%%%%%%

%%%%%%%%%% Insert the texts which can accomdate on firstpage in the tag "fmtext" %%%%%

%\begin{fmtext}

%\end{fmtext}

%%%%%%%%%%%%%%% End of first page %%%%%%%%%%%%%%%%%%%%%

\maketitle

\section{Introduction}
Computational Fluid Dynamics (CFD) allows new aeromechanical designs to be evaluated by modelling the flow around the proposed geometry from first principles. In contrast to simpler correlation based design methods, this has allowed designers to explore the design space beyond the existing design envelope \cite{Michelassi2021TuboReview}. While this has presented opportunities, the Reynolds-averaged Navier–Stokes (RANS) models, which are typically employed by industry, suffer from fundamental weaknesses that undermine their efficacy as a design tool \cite{ijtpp7020016}. Within RANS, the intermediate and small energy scales are modelled rather than resolved directly. This has the effect of making the model inexpensive compared to Large Eddy Simulations (LES) and Direct Numerical Simulations (DNS), where comparatively smaller scales are resolved, but reduces the accuracy of the model (see, e.g. \cite{chaouat2017state}). %In practice, safety margins are placed on RANS predictions to compensate for this.

Lately, attention has been paid to employing machine learning (ML) methods to leverage high-fidelity data from LES, DNS, and experiments, improving the accuracy of RANS models while adding minimal cost to the computation \cite{duraisamy2018turbulence}. In many ways this is a natural problem for the application of machine learning due to the large number of parameters and non-linear nature of the required model \cite{2021AcMSn..37.1718B}. Two broad strategies can be identified within these works: one approach is to learn a modified Reynolds stress tensor directly from the high-fidelity data, concentrating on regions where certain assumptions of the RANS results break down. This could be done through a neural network, for instance Frey et al. \cite{Frey2021DNN} and Ling \cite{Ling2016DNN} used neural networks trained on DNS data to correct the eddy viscosity term in the Boussinesq equation. Weatheritt and Sandberg \cite{Weatheritt2016Evolutionary} employed Gene Expression Programming (GEP), in which the explicit algebraic Reynolds stress (EASM) framework of Pope \cite{Pope1975GeneralEffVisc} was used to perform a symbolic regression to learn improved nonlinear expressions for the anisotropy. The explicit expressions that this approach provides are more interpretable than a neural network, but this is at the expense of degrees of freedom in the model. 

An alternative strategy is to modify the RANS turbulence equations to include a spatial correction function allowing the solution to better fit a reference mean flow solution. If data is available for the entire flow field a frozen approach can be taken where the variables to be fitted are kept constant (e.g. velocity and $k$) while solving for the remaining RANS modelling variables and corrections needed. This was the approach taken by Schmelzer et al. \cite{Schmelzer2020SpaRTA}. If only sparse data is available a field inversion or data assimilation technique can be used. An objective function composing of the error between the reference and RANS data is minimised, typically using adjoint methods to obtain the objective function gradients. The correction function for the entire flow field can be obtained even if only lines or planes of data \cite{he_data}, surface pressure coefficient or lift coefficient \cite{Singh2017MLAirfoil} are available. A supervised learning algorithm is then used to learn the correction function from the local flow variables. Examples of this approach include Parish and Duraisamy \cite{Parish2016FieldInv} and more recently, Ho and West \cite{ho2021field}. 

Research employing both strategies has yielded promising results but has also provoked questions as to the practical usage of machine learning methods to improve turbulence modelling \cite{gen1, gen2, gen3}. Three such challenges, that are the subject of this paper, include: 

\begin{itemize}
    \item How far should a machine learning algorithm's predictions be trusted when presented with new geometries and flow conditions?
    \item How can we determine if a machine learning corrected turbulence model is extrapolating beyond its training dataset?
    \item Can the model handle large, heterogeneous datasets? 
\end{itemize}

To address these challenges, this paper couples a framework for probabilistic machine learning with field inversion. Training an ensemble of models separately provides a modular solution that can naturally handle heterogeneous data. By analysing the variance of the machine learning correction, we show that the framework can improve the accuracy of RANS simulations for geometries similar to those in the training data while simultaneously, and perhaps more significantly, rejecting the machine learning correction for geometries beyond the range of the training data. 

This paper is structured as follows: in Section 2 we discuss the field inversion method, the identification of the non-dimensional flow features, and the probabilistic machine learning model. Section 3 introduces the training cases that were used to generate the heterogeneous data and compares the results of the flow inversion with the reference solutions. The training of the probabilistic machine learning framework on this data and the methods used as a baseline for its performance is also discussed. Finally, we demonstrate the application of the framework to flows around four previously unseen geometries.  

%\begin{itemize}
%\item Deficiencies in RANS models (see chat from Vittorio in \cite{ijtpp7020016})

%\item Interest in using data from LES/DNS/experiments to enrich the RANS model such as through NNs, GEP

%\item Handling uncertainty? Zabaras paper \cite{GENEVA}. Handling unseen geometry? Comment on implications for risk averse industries like aeronautics

%\item Data from multiple geometries? Heterogeneous data sources? Discussion of Simpson's paradox? 
%\end{itemize}

\section{Methodology} \label{sec:methodology}

\subsection{Field Inversion}
The specific dissipation rate, $\omega$, equation of the $k-\omega\ SST$ model was modified to allow the RANS solution to adapt to the reference solution. A spatially varying correction factor, $\beta_c(\boldsymbol{x})$, was applied to the production term, resulting in the modified $\omega$ equation:
\begin{align}
    \frac{D \omega}{Dt}=-\nabla \cdot (\rho D_\omega \nabla \omega)+\beta_c(\boldsymbol{x})\frac{\rho \gamma G}{\epsilon}-\frac{2}{3}\rho \gamma \omega (\nabla \cdot \boldsymbol{u})-\rho \beta \omega^2-\rho(F_1-1)CD_{k \omega}.
\end{align}
Details of the various terms of the turbulence model can be found in the reference from Menter \cite{menter_kwsst_2003}. Matai and Durbin \cite{matai_durbin_2019} have studied the effects of placing the correction factor on different parts of the $\omega$ equation and found that the exact $\nu_t$ field was recovered whether $\beta_c$ was applied as a multiplier to the production or destruction term. Duraisamy et al. \cite{Duraisamy2015Inverse} explains that introducing the correction factor as a multiplication to existing terms is equivalent to adding a new correction term to the model. The advantage of having $\beta_c$ as a multiplication is that it is dimensionless and the optimised value would be in the same order of magnitude as the uncorrected value of $1$.

Standard RANS turbulence models are known to perform badly in separated flows, typically overpredicting the separation region with delayed reattachment. This is due to models underestimating the larger eddies present in the shear layer of separated flows, leading to an underprediction of turbulent shear stresses and reduced turbulent mixing in the separated region (see, e.g. \cite{rumsey_flow_control}). Introducing $\beta_c$ into the $\omega$ equation changes the balance of turbulent production to dissipation locally, allowing the model to overcome the described shortcomings. This is reflected in the inverted results shown in Section \ref{sec:data_gen_training} where the corrections for all cases are most active in the shear layer of the separation region, mostly with $\beta_c<1$ causing an increase in $k$.

Given an uncorrected solution of the $k-\omega$ SST model and a high-fidelity or experimental reference solution, we wish to obtain the estimate of the field $\beta_c(\boldsymbol{x})$ that minimises the discrepancy between the two flow fields. The required $\beta_c$ field is determined by discretising the flow field into $n$ cells and solving the optimisation problem:   

\begin{align}
    &\underset{\boldsymbol{\beta_c}}{\textrm{min}} \;
    \mathcal{J}(\boldsymbol{\beta_c}, \boldsymbol{w})         \label{eq:obj_fun1}\\ 
    &\text{such that} \;\mathcal{N}(\boldsymbol{\beta_c}, \boldsymbol{w})=0,\nonumber
\end{align}
where $\boldsymbol{\beta_c}\in\mathbb{R}^n$ collects the correction factor at every cell and $\boldsymbol{w}$ represents the conserved variables. $\mathcal{N}(\cdot)$ is the nonlinear discretisation operator that solves the flow equations and $\mathcal{J}(\cdot)$ an objective function that computes the discrepancy between the RANS solution and reference solution on $n_a$ cells (which can be less than the total number of cells in the simulation, $n$). The objective function is defined as:

\begin{align}
    \mathcal{J}(\boldsymbol{\beta_c}, \boldsymbol{w})=\frac{1}{n_a}\sum_{i=1}^{n_a}\Big( \boldsymbol{u}_i(\boldsymbol{\beta}_{\boldsymbol{c}}, \boldsymbol{w})-\boldsymbol{u}_{\text{ref}, i}\Big)^2+\frac{\lambda}{n}\sum_{i=1}^n (\boldsymbol{\beta}_{\boldsymbol{c},i}-1)^2,
    \label{eq:obj_fun2}
\end{align}
where $\boldsymbol{u}_i(\boldsymbol{\beta_c}, \boldsymbol{w})$ refers to the solved RANS velocity field at the $i$\textsuperscript{th} cell and $\boldsymbol{u}_{\text{ref}, i}$ the reference solution at that cell. The first term represents the $L_2$ distance between the solved RANS velocity field and the reference solution, while the second term is a regularisation term for the correction factor, with $\lambda$ a regularisation constant. The optimisation problem described in Equations \eqref{eq:obj_fun1} and \eqref{eq:obj_fun2} was solved using a gradient descent algorithm. At the end of the optimisation, a $\beta_c$ field that is piecewise constant within each cell is obtained. A discrete adjoint formulation was employed to obtain the required gradients: 

\begin{align}
    \Bigg(\frac{\partial R}{\partial \boldsymbol{w}}\Bigg)^\top\boldsymbol{\phi}=\frac{\partial \mathcal{J}}{\partial \boldsymbol{w}}, \\
    \frac{d\mathcal{J}}{d\boldsymbol{\beta_c}}=\frac{\partial\mathcal{J}}{\partial \boldsymbol{\beta_c}}-\boldsymbol{\phi}^\top\frac{\partial R}{\partial \boldsymbol{\beta_c}}, \nonumber
\end{align}

where $R\in\mathbb{R}^{(n\times n_w)}$ represents the residuals being driven to zero by the flow solver, $\boldsymbol{w}$ the conserved variables, $n_w$ the number of conserved variables, and $\boldsymbol{\phi}\in\mathbb{R}^{(n\times n_w)}$ is the adjoint solution. Recall that $\boldsymbol{\beta_c}\in\mathbb{R}^n$ refers to the correction at every cell.

A finite volume solver with second order upwind discretisation was used in this work. The incompressible RANS equation with a linear eddy viscosity model was solved using the SIMPLE algorithm. Once a converged primal flow field is obtained, a discrete adjoint solver is used to calculate the gradients for the optimisation. 

\subsection{Feature space representation}

Having performed the inversion, $\beta_c$ is regressed to a non-dimensional feature space that describes the flow, denoted $\boldsymbol{\alpha}\in\mathbb{R}^{52}$. Five of these features were manually engineered, namely:

\quad Streamline pressure gradient:
\begin{align}
    \frac{2}{\pi}\textrm{arccos}\bigg(\frac{\nabla p\cdot \boldsymbol{u}}{\sqrt{||\nabla p||^2||\boldsymbol{u}||^2}}\bigg)-1.
    \label{eq:feat_manual_1}
\end{align}

\quad Turbulence time scale ratio:
\begin{align}
    2\bigg[\bigg(\frac{S}{S+\omega}\bigg)^{0.25}-0.5\bigg].
\end{align}

\quad Q-criterion:
\begin{align}
    \frac{S^2-\Omega^2}{S^2+\Omega^2}. 
\end{align}

\quad Turbulence intensity:
\begin{align}
    2\bigg[\bigg(\frac{k}{0.5||\boldsymbol{U}||^2+k}\bigg)^{0.25}-0.5 \bigg].
    \label{eq:feat_manual_4}
\end{align}

\quad Turbulent viscosity ($\nu_t$) ratio:
\begin{align}
    2\bigg[\bigg(\frac{\nu_t}{100\nu+\nu_t}\bigg)-0.5\bigg]. 
\end{align}

$S$ and $\Omega$ represent the Frobenius norm of the strain-rate and rate of rotation tensor respectively (i.e. $S=\sqrt{\sum_i\sum_jS_{ij}S_{ij}}$). $p$ represents the kinematic pressure, $k$ the turbulent kinetic energy, $\boldsymbol{U}$ the Reynolds averaged velocity, and $\nu$ the kinematic viscosity. The exponent of 0.25 was applied to the turbulence time scale ratio and turbulence intensity features to improve the distribution. It is noted that equations \eqref{eq:feat_manual_1} and \eqref{eq:feat_manual_4} are not Galilean invariant, however, the test cases in this work share the same frame, hence the effect is not significant.

The other 47 features were obtained based on the minimal integrity basis for rotational invariance of the tensor set $\{\hat{S}, \hat{\Omega}, \hat{A_p}, \hat{A_k}\}$, with the overhats denoting non-dimensionalisation. The tensors $A_p$ and $A_k$ are the pressure and turbulent kinetic energy gradients mapped to an antisymmetric tensor through $-I\times\nabla p$ for example. More details on these features may be found in Wu et al. \cite{Wu2018ML} and Ling et al. \cite{LING201622}. %Together, these 52 non-dimensional features formed a feature space in which each cell in the flow field could be characterized, denoted $\boldsymbol{\alpha}\in\mathbb^{52}$. 

%%%%%%%%%%%%%%%%%%%%%%%%%%%%%%%%
\subsection{Probabilistic Machine Learning Framework}
Having obtained the non-dimensional features, $\boldsymbol{\alpha}$, from the uncorrected solution to the RANS equations and the correction factor, $\beta_c$, from field inversion, a probabilistic machine learning framework is introduced to learn the mapping $\boldsymbol{\alpha}\rightarrow \beta_c$. Recognising that in practice reference data is likely to come from heterogeneous data sources (for instance from PIV, DNS, and LES) and a range of geometries, there is a strong likelihood that this will be a one-to-many mapping. A machine learning model trained on the entire dataset could fail to learn the trends within the individual datasets, limiting its predictive power (Simpson's paradox \cite{pml1Book}). To address this, we introduce a probabilistic machine learning model for the mapping based on an ensemble of Gaussian Process Emulators (GPEs).

\smallskip
Given a training dataset of $n_t$ parameter points $X=[\boldsymbol{\alpha}^{(1)}, \dots, \boldsymbol{\alpha}^{(n_t)}]$ and $Y=[\beta_c^{(1)},\dots, \beta_c^{(n_t)}]$, from $n_m$ separate data sources, an interpolation function, $\Psi$, is developed for each data source, i.e.:

\begin{align}
    {{\beta}_c}(\boldsymbol{{\alpha}})={\Psi_k}(\boldsymbol{{\alpha}})+\boldsymbol{\epsilon}_k,\; \; k=1,\dots, n_m,
\end{align}
where $\boldsymbol{\epsilon}\sim\mathcal{N}(\boldsymbol{\mu_0},\text{diag}(\boldsymbol{\sigma}_{{\epsilon}}))$ is assumed to be independent, identically distributed Gaussian noise with variance $\boldsymbol{\sigma}^2_{\epsilon}\in\mathbb{R}^{n_m}$ and mean $\boldsymbol{\mu_0}\in\mathbb{R}^{n_m}$. $\Psi_k$ refers to the interpolation function for the $k$\textsuperscript{th} data source. A Gaussian Process prior is assumed for these functions, allowing the posterior density for $\beta_c$ at a new parameter point in feature space, $\boldsymbol{\alpha^*}$, to be expressed as a Gaussian distribution: 

\begin{align}
    {\beta_c}|X_k,Y_k,\boldsymbol{\alpha^*}\sim {N}({\mu}_k,{\sigma}_k),%\otimes\hat{\Omega}),
\end{align}
where $X_k$ and $Y_k$ denote the subset of the training data originating from the $k$\textsuperscript{th} high-fidelity dataset, allowing different sets of data such as discrete measurements, high-fidelity CFD, and potentially experience based rules to be included within the framework. Assuming a constant mean function, $\boldsymbol{\mu_0}$, the mean and standard deviation of the $k$\textsuperscript{th} GPE are denoted: 
\begin{align}
 {\mu}_k&={\boldsymbol{\mu}}_{\boldsymbol{0},k}+K(\boldsymbol{\alpha^*},X)^\top K(X,X)^{-1}(Y_k-{\boldsymbol{\mu_0}}_{,k}), \label{eq:gp_moms} \\
 {\sigma}^2_k&=K(\boldsymbol{\alpha^*},\boldsymbol{\alpha^*})-K(\boldsymbol{\alpha^*},X)^\top K(X,X)^{-1}K(\boldsymbol{\alpha^*},X), \nonumber
\end{align}
where $K$ refers to the covariance matrix, with elements given by:
\begin{align}
    K_{ij}(X,X)=k(\boldsymbol{\alpha}^{(i)}, \boldsymbol{\alpha}^{(j)})+\sigma^2_{\epsilon_k}\delta_{ij}. 
\end{align}

$k(.)$ represents the covariance function and $\delta$ the Kronecker delta. In this work squared exponential kernels were used for the covariance function. The hyperparameters for the GPE are inferred from the data, at a cost $O(n_t^3)$. To mitigate this cost when $n_t$ is large (e.g. $n_t>10^3$), we use a stochastic variational regression for the GP hyperparameters. We refer the interested reader to Blei et al. \cite{Blei_2017} and Hensman et al. \cite{hensman} for more details on this and to the canonical reference books of Rasmussen and Williams \cite{GP1} and Santner et al. \cite{GP3} for more details on GPEs in general.

Having obtained the ensemble of trained GPEs ${N}(\mu_k(\boldsymbol{\alpha}), \sigma_k(\boldsymbol{\alpha})),\;k=1,\dots,n_m$, we discuss how their contributions are weighted within the probabilistic machine learning framework and the criterion by which this correction factor is accepted. For the unseen parameter point in feature space, $\boldsymbol{\alpha^*}$, the mean and standard deviation for each GPE in the ensemble is determined through \eqref{eq:gp_moms}, with the mean and variance of the ensemble obtained through the weighted summations:

\begin{align}
    \mu^*(\boldsymbol{\alpha^*})&=\sum_{k=1}^{n_m} w_k\mu_k(\boldsymbol{\alpha^*}),\\
    \sigma^{*2}(\boldsymbol{\alpha^*})
    &= {{\sigma^*}_\mu^2(\boldsymbol{\alpha^*})+\sigma^{*2}_\sigma(\boldsymbol{\alpha^*})}, \nonumber 
\end{align}
where the standard deviation of the ensemble includes contributions from the variance of the individual GPEs, $\sigma^*_\sigma$, but also the spread of their means, $\sigma^*_\mu$, as in Lakshminarayanan et al. \cite{ensemble}, with:
\begin{align}
    \sigma^{*2}_\mu&= {\sum_{k=1}^{n_m} w_k \mu^2_k(\boldsymbol{\alpha^*}})-\mu^{*2}(\boldsymbol{\alpha^*}), \\
    \sigma^{*2}_\sigma&= \sum_{k=1}^{n_m} w_k\sigma_k^2(\boldsymbol{\alpha^*}). \nonumber
\end{align}
The individual mixture weights satisfy $\sum_{k=1}^{n_m}w_k=1$ and are determined using the inverse weighted variance:
\begin{align}
    w_k=\frac{{1}/{\sigma_k^2}}{\sum_{j=1}^{n_m}{{1}/{\sigma_j^2}}}.
\end{align}
More confident GPEs with lower variance are therefore assigned more weight in this framework. A key motivation for employing probabilistic machine learning is that a measure of the uncertainty accompanies the prediction of the correction factor in the form of the variance. The variance is used to identify two scenarios: where the framework is generalising to a region of feature space that is outside of the training data; and secondly where $\boldsymbol{\alpha^*}$ is within the training data but there is significant disagreement between the predictions of the GPEs. As will be seen, in such cases applying the data-driven correction could worsen the accuracy of RANS compared to the uncorrected solution. Consequently, we apply the following acceptance criterion to the framework's predictions: 
\begin{align}
\beta_c=
    \begin{cases}
        \mu^*(\boldsymbol{\alpha^*})\; \textrm{if} \; \sigma^*(\boldsymbol{\alpha^*})\leq \bar{\sigma} \\
        1\; \text{otherwise}
    \end{cases},
    \label{eq:crit}
\end{align}
where $\bar{\sigma}$ is a user specified tolerance that must be chosen to balance the magnitude of the correction against conservatism of the approach when applied to flow conditions or geometries that are outside of the training data. In addition to the described ensemble of GPEs, a separate deep ensembles (DEs) model based on Lakshminarayanan et al. \cite{ensemble} was trained with a similar acceptance criterion to provide a comparison to the proposed method. Both models were trained on a training dataset harvested from three experimental and high-fidelity datasets that are described in the next section. In a departure from the approach commonly taken in the literature, where uncertainty is minimised in the freestream and maximised in regions of the flow with high Reynolds stress anisotropy (see, e.g. \cite{ling2015evaluation}), the predictive uncertainty of the model is instead minimised in regions where $\beta_c$ is active. This is achieved by removing points from the training dataset in the interval $\beta_c \in[0.9, 1.1]$, recalling that $\beta_c=1$ indicates that the correction is inactive. 

Once the models have been trained, the predictions on future cases are added as a one-time-correction to a converged uncorrected flow field which is solved again to obtained the final flow field. The probabilistic machine learning framework described here is not limited to such an implementation and can also be applied to an embedded ML model that is called every iteration without any loss in generality.

%\textcolor{red}{Different tolerance to GPE, how we prepare the training data ([0.9,1.1])}

%Correction to RANS is both a function of the mean and variance of the mixture model at the parameter point $\boldsymbol{x^*}$:

%\begin{align}
%    \hat{\beta}=1+(\mu^*-1)g(\sigma^*)
%\end{align}
%where $g(\cdot)$ has the following properties:
%\begin{itemize}
%    \item Is monotonic 
%    \item $g(\sigma^*)\rightarrow 1$ as $\sigma^*\rightarrow 0$
%    \item $g(\sigma^*)\rightarrow 0$ as $\sigma^*\rightarrow \infty$
%\end{itemize}

%Practical effects: the more confident the mixture (i.e. $\sigma^*\rightarrow 0$) then $\hat{\beta}\rightarrow\mu^*$. Conversely, as the uncertainty of the mixture model grows, it's contribution becomes negligible and RANS is used without a ML correction i.e. $\hat{\beta}\rightarrow 1$ as $\sigma^* \rightarrow \infty$. $g(\cdot)$ used in this paper is a modification of the exponential linear unit:
%\begin{align}
%    g(\sigma^*)=    \begin{cases}
%        1\; &\text{if} \,\sigma^*\leq c_2 \\
%        \text{exp}(-c_1 (\sigma^*-c_2))\; &\text{if}\,\sigma^*>c_2 \\
%    \end{cases},
%\end{align}

%$c_1$ and $c_2$ are constants that are determined by optimisation (similar idea to NOMU). 

\section{Data generation and model training} \label{sec:data_gen_training}
The training data was collected from a range of openly available experimental and high fidelity results. The three cases chosen were: the NASA wall mounted hump from Greenblatt et al. \cite{greenblatt_exp_nasa_hump_2006}, curved backward facing step from Bentaleb et al. \cite{lesRoundedStep} and cylinder from Lehmkuhl et al. \cite{Lehmkuhl_2011}. The data was a mix of types which included PIV, LES, and DNS respectively. One common feature of these cases is flow separation from a curved surface in adverse pressure gradient. 

\subsection{NASA hump training case}
A modified Glauert hump geometry was chosen in in the original experiment as it featured a separation point not fixed by geometry that is representative of an aerodynamic body. After a relatively long forebody, the hump encounters a relatively short concave ramp where the flow separates. The separated flowfield was reported to be insensitive in a large range of Reynolds number ($2.4\times 10^6<Re<26\times 10^6$). A series of experiments with active blowing just before the ramp was performed originally but only the baseline results without flow control was used in this work. Compared to the other computational data sets used in this work, the Reynolds number here is higher at $Re_c=9.36\times 10^5$.

\begin{figure}[h]
    \centering
    \includegraphics[width=\textwidth]{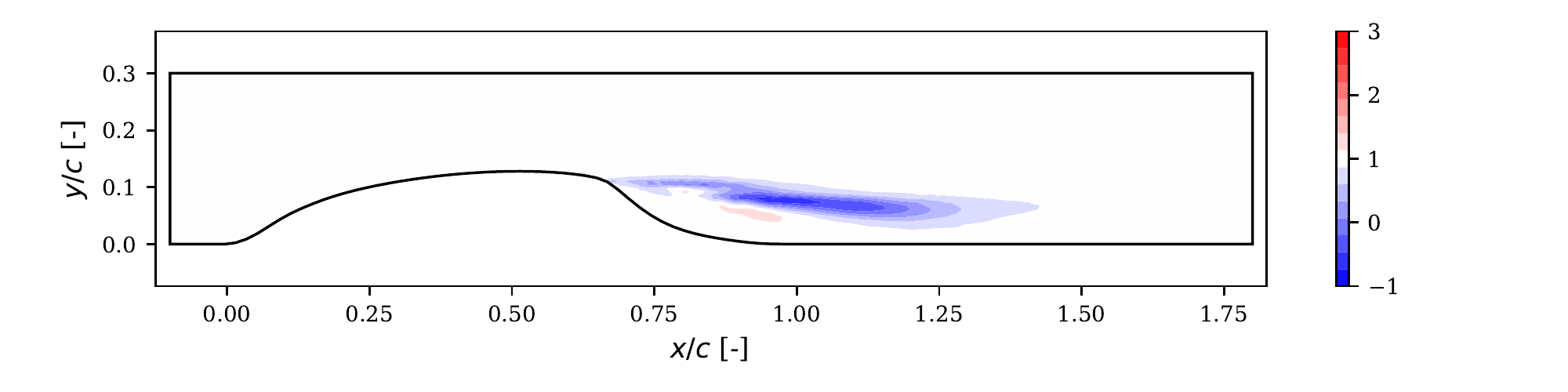}
    \caption{Inverted $\beta_c$ field for the NASA hump.}
    \label{fig:nasa_hump_beta}
\end{figure}

\begin{figure}[h]
    \centering
    \begin{subfigure}[b]{0.32\textwidth}
        \centering
        \includegraphics[width=\textwidth]{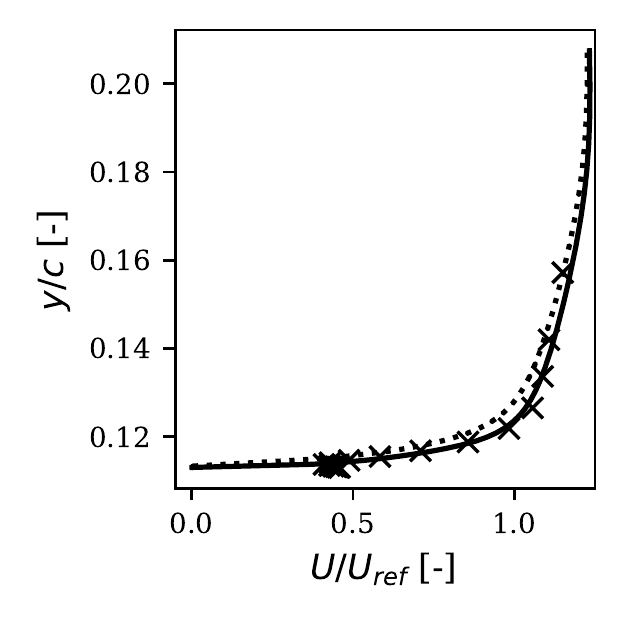}
        \caption{$x/c=0.66$.}
        \label{fig:nasa_hump_vel_0.66}
    \end{subfigure}
    \begin{subfigure}[b]{0.32\textwidth}
        \centering
        \includegraphics[width=\textwidth]{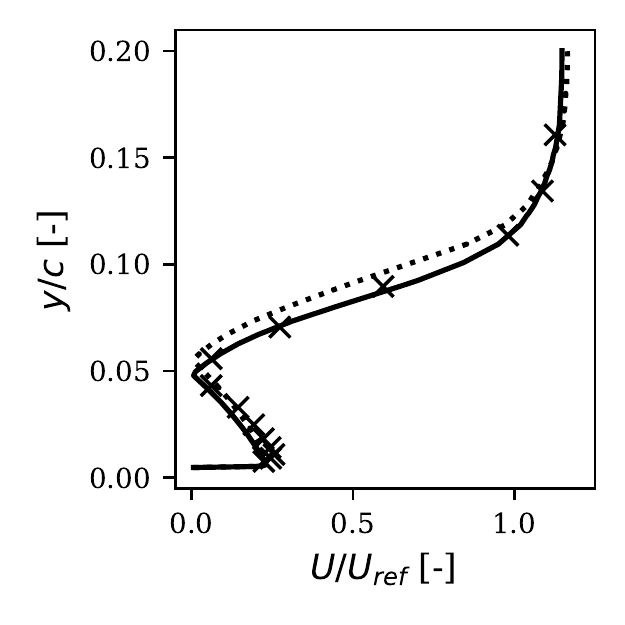}
        \caption{$x/c=0.90$.}
        \label{fig:nasa_hump_vel_0.90}
    \end{subfigure}
    \begin{subfigure}[b]{0.32\textwidth}
        \centering
        \includegraphics[width=\textwidth]{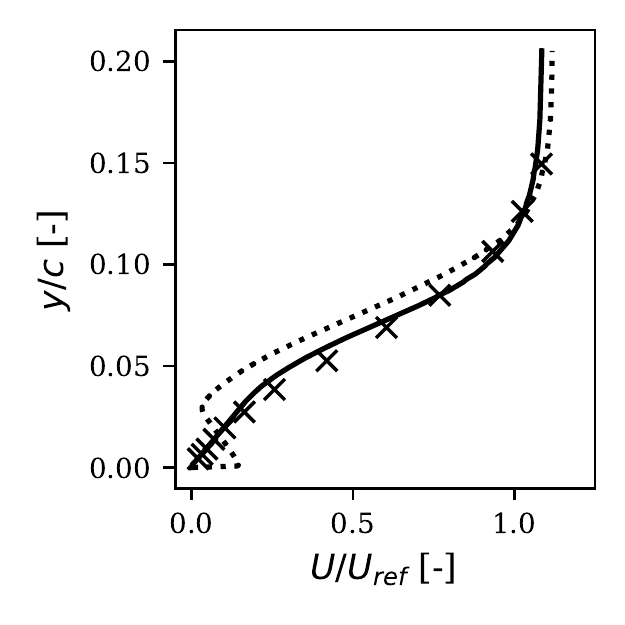}
        \caption{$x/c=1.10$.}
        \label{fig:nasa_hump_vel_1.10}
    \end{subfigure}
    \caption{Velocity profiles of inverted NASA hump case. Reference PIV (\markercross), uncorrected (\dotted), inverted (\full).}
    \label{fig:nasa_hump}
\end{figure}

The PIV data was assimilated as lines of data between $0.8\leq x/c\leq 1.3$ with spacings of $0.1$. Two additional lines of data at $x/c=0.65$ and $0.66$, just after the peak of the hump, was also included. The optimisation was stopped after 63 iterations when the objective function reduced an order of magnitude and started to plateau. The resultant correction field was most active in the separated region away from the hump (Figure \ref{fig:nasa_hump_beta}). The optimised $\beta_c$ was mostly less than $1$ with a small region of increased $\beta_c$. Figure \ref{fig:nasa_hump} shows good agreement between the inverted and measured velocity fields.

\subsection{Curved backward facing step training case}
Similar to the NASA hump case, the curved backward facing step does not have a fixed separation point. One difference here is that there is no upstream acceleration of the flow, hence it features flow separation from a fully developed turbulent boundary layer. Data was taken from averaged LES velocities with the original simulation performed at height based Reynolds number $Re_h=13700$. 

\begin{figure}[h]
    \centering
    \includegraphics[width=0.9\textwidth]{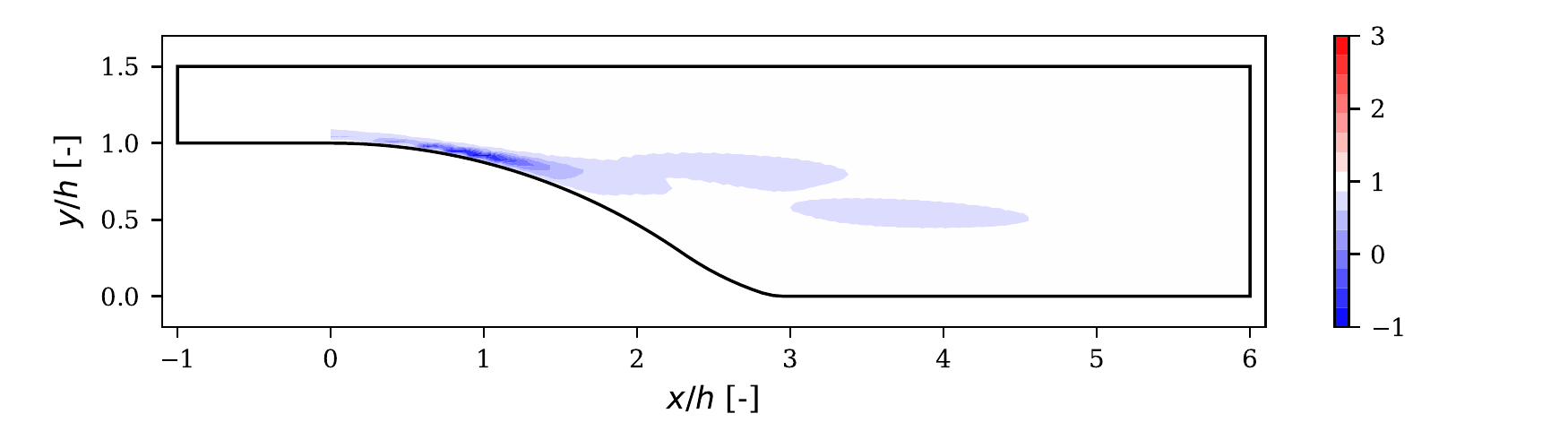}
    \caption{Inverted $\beta_c$ field for the curved backward facing step.}
    \label{fig:bfs_beta}
\end{figure}

\begin{figure}[h]
    \centering
    \begin{subfigure}[b]{0.32\textwidth}
        \centering
        \includegraphics[width=\textwidth]{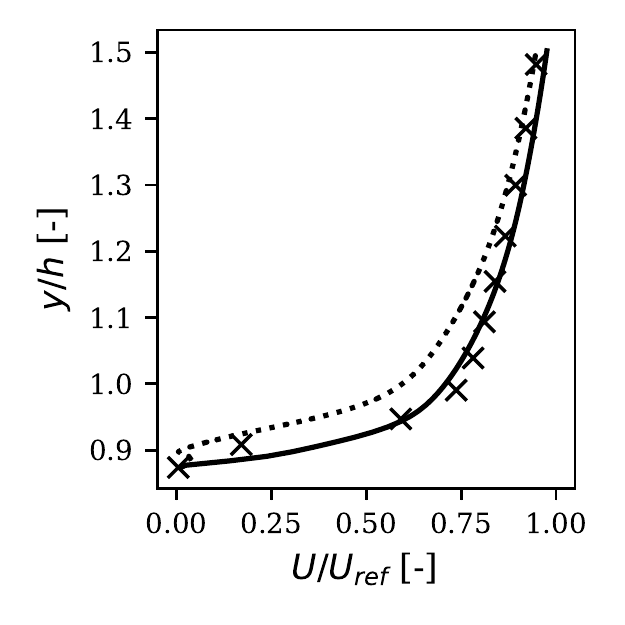}
        \caption{$x/h=1$.}
        \label{fig:bfs_vel_1}
    \end{subfigure}
    \begin{subfigure}[b]{0.32\textwidth}
        \centering
        \includegraphics[width=\textwidth]{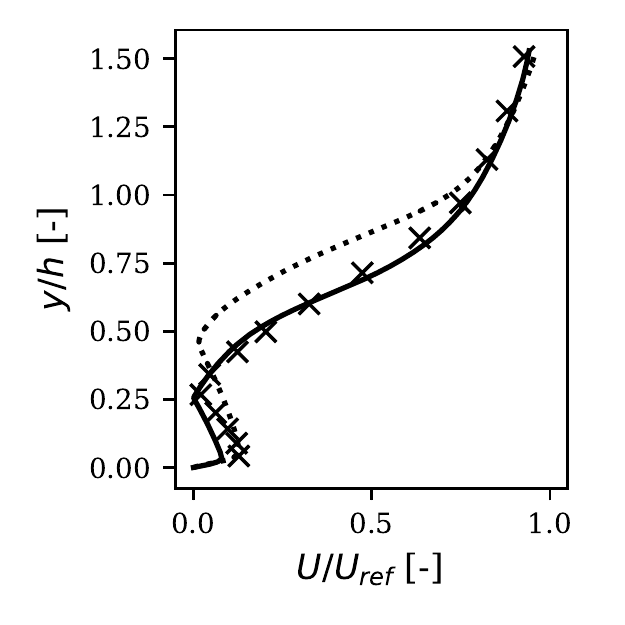}
        \caption{$x/h=3$.}
        \label{fig:bfs_vel_3}
    \end{subfigure}
    \begin{subfigure}[b]{0.32\textwidth}
        \centering
        \includegraphics[width=\textwidth]{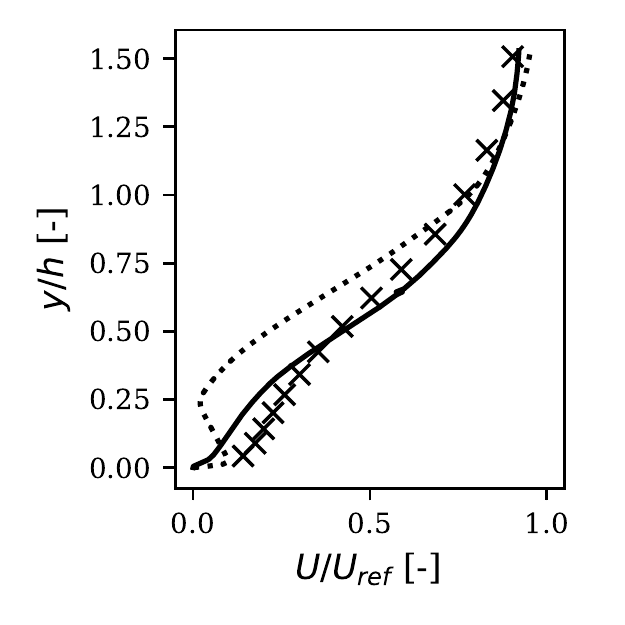}
        \caption{$x/h=5$.}
        \label{fig:bfs_vel_5}
        \end{subfigure}
    \caption{Velocity profiles of curved backward facing step. Reference LES (\markercross), uncorrected (\dotted), inverted (\full).}
    \label{fig:bfs}
\end{figure}

Data was assimilated in a 2D region where $0<x/h<8$ and $0<y/h<2$ and the optimisation was stopped after 48 iterations when the objective function plateaued after dropping an order of magnitude. Figure \ref{fig:bfs_beta} shows the inverted $\beta_c$ field which had higher corrections in the boundary layer compared to the other inverted cases. As a result, the momentum in the boundary layer was increased at $x/c=0.66$, better matching the LES velocity profile as shown in Figure \ref{fig:bfs_vel_1}. The velocity profiles in the separation region also matches well with the reference LES results (Figure \ref{fig:bfs}).

\subsection{Cylinder training case}
The reference data for this case was based on a relatively long DNS simulation where the low frequency unsteadiness of the separation bubble was the subject of interest. This case features separated shear layers and its interaction with the turbulent wake. The diameter based Reynolds number is $3900$. 

\begin{figure}[h]
    \centering
    \includegraphics[width=0.9\textwidth]{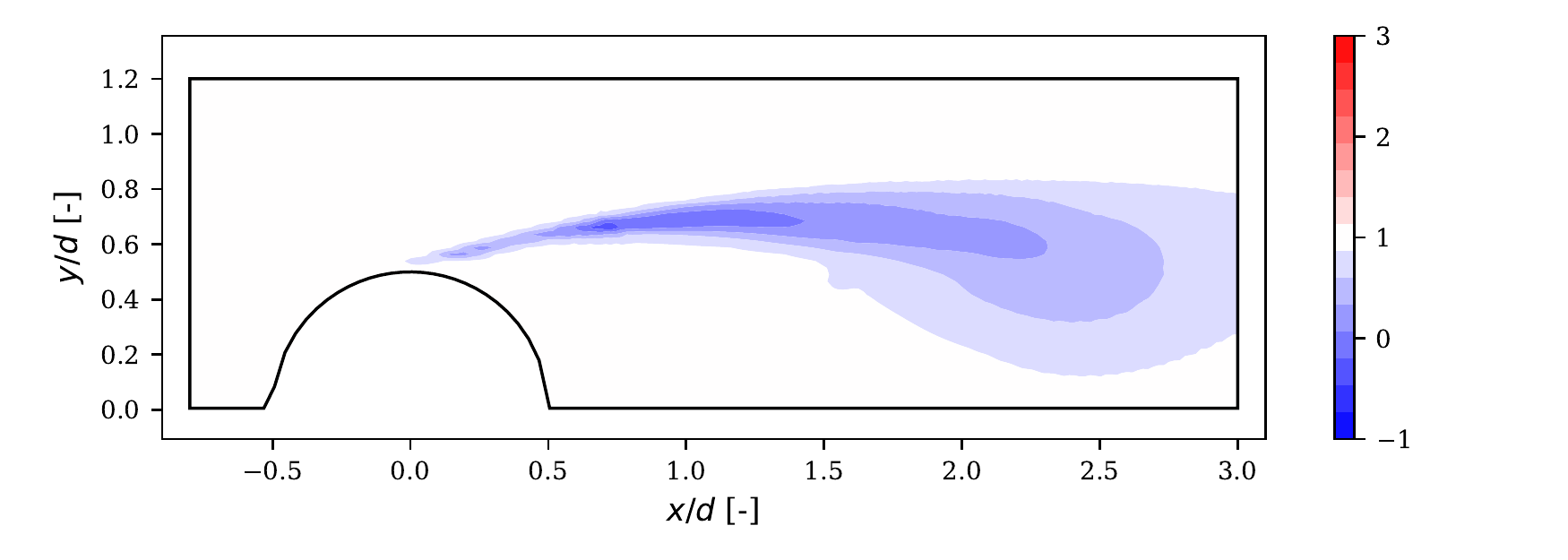}
    \caption{Inverted $\beta_c$ field for the cylinder.}
    \label{fig:cyl_beta}
\end{figure}

\begin{figure}[h]
    \centering
    \begin{subfigure}[b]{0.32\textwidth}
        \centering
        \includegraphics[width=\textwidth]{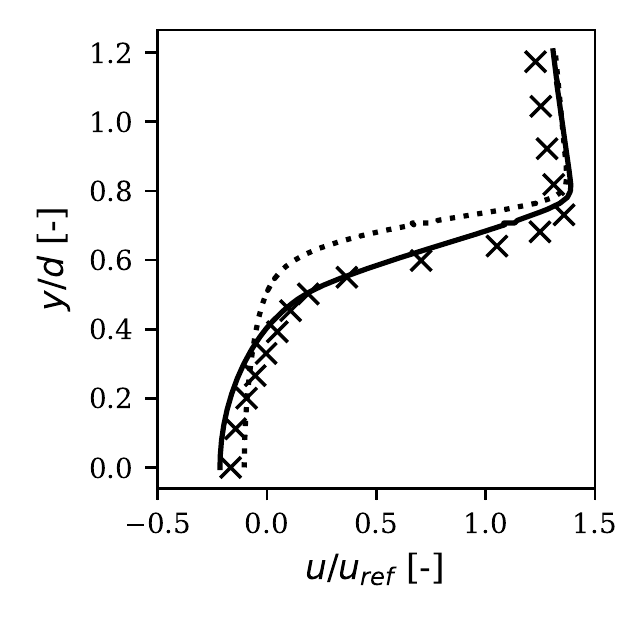}
        \caption{$x/d=1.06$.}
        \label{fig:cyl_vel_1.06}
    \end{subfigure}
    \begin{subfigure}[b]{0.32\textwidth}
        \centering
        \includegraphics[width=\textwidth]{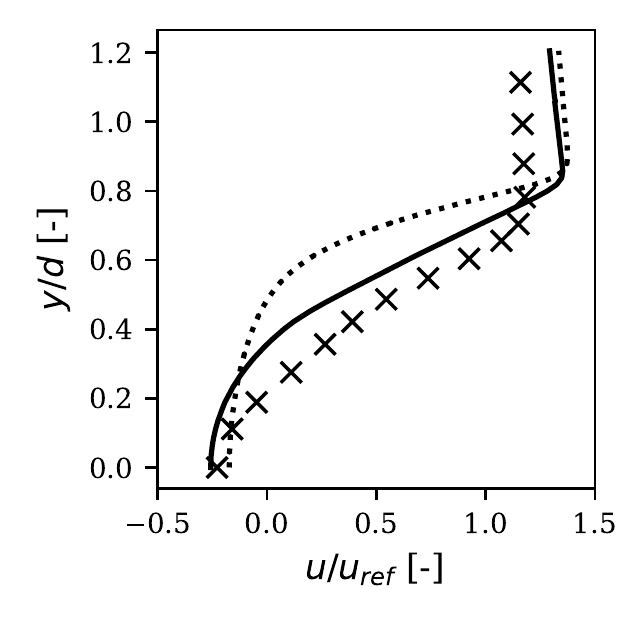}
        \caption{$x/d=1.54$.}
        \label{fig:cyl_vel_1.54}
    \end{subfigure}
    \begin{subfigure}[b]{0.32\textwidth}
        \centering
        \includegraphics[width=\textwidth]{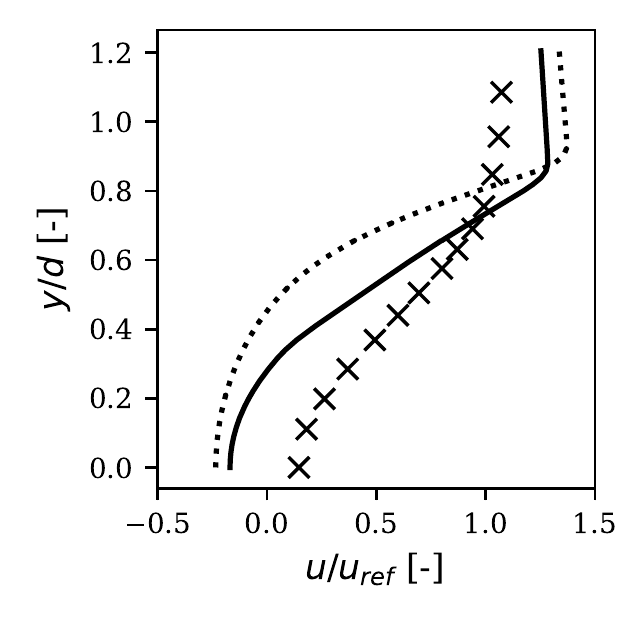}
        \caption{$x/d=2.02$.}
        \label{fig:cyl_vel_2.02}
    \end{subfigure}
    \caption{Velocity profiles of cylinder in cross flow. Reference DNS (\markercross), uncorrected (\dotted), inverted (\full).}
    \label{fig:cylinder}
\end{figure}

The data in the near wake on lines at $x/c=1.06$, $1,54$ and $2.02$ was assimilated. The optimisation was stopped after 80 iterations when the objective function started to plateau. Compared to the other inverted results there was a larger region of correction in the wake towards the end of the separation bubble. This could be due to the poor prediction of the relaxation region post separation of the uncorrected results as shown in the velocity profiles in Figure \ref{fig:cylinder}. The velocity profiles between the inverted and DNS flow field at $x/c=1.06$ and $x/c=1.54$ showed good agreement and the inverted flow field provided some improvements at $x/c=2.02$. There was an overall increase in the diffusion of the wake in the inverted flow field.

\FloatBarrier
\subsection{Training the ensemble models}
\begin{figure}[h]
    \centering
    \begin{subfigure}[b]{0.32\textwidth}
        \centering
        \includegraphics[width=\textwidth]{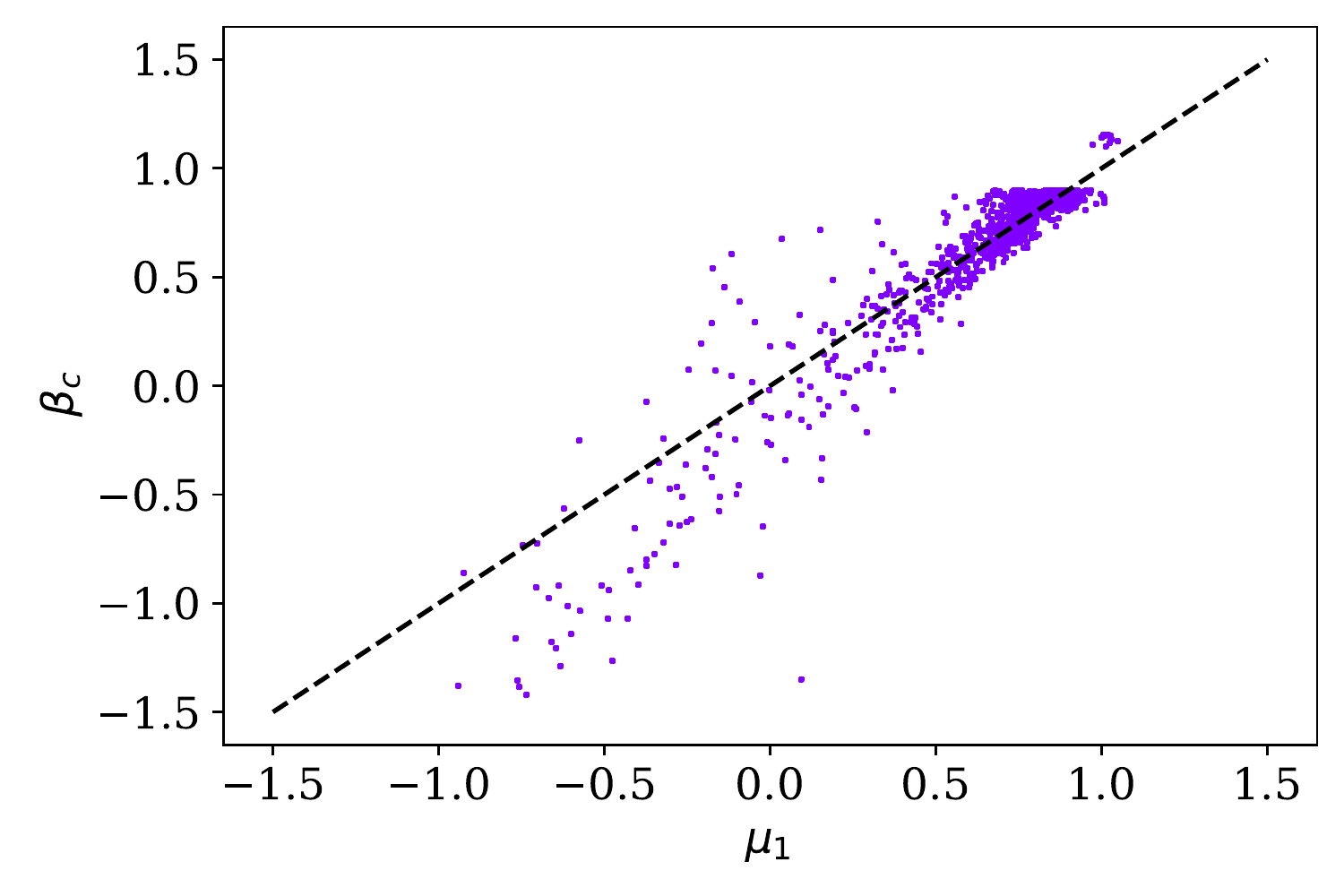}
        \caption{Curved Backwards facing step}
        \label{fig:bbfs_train}
    \end{subfigure}
    \begin{subfigure}[b]{0.32\textwidth}
        \centering
        \includegraphics[width=\textwidth]{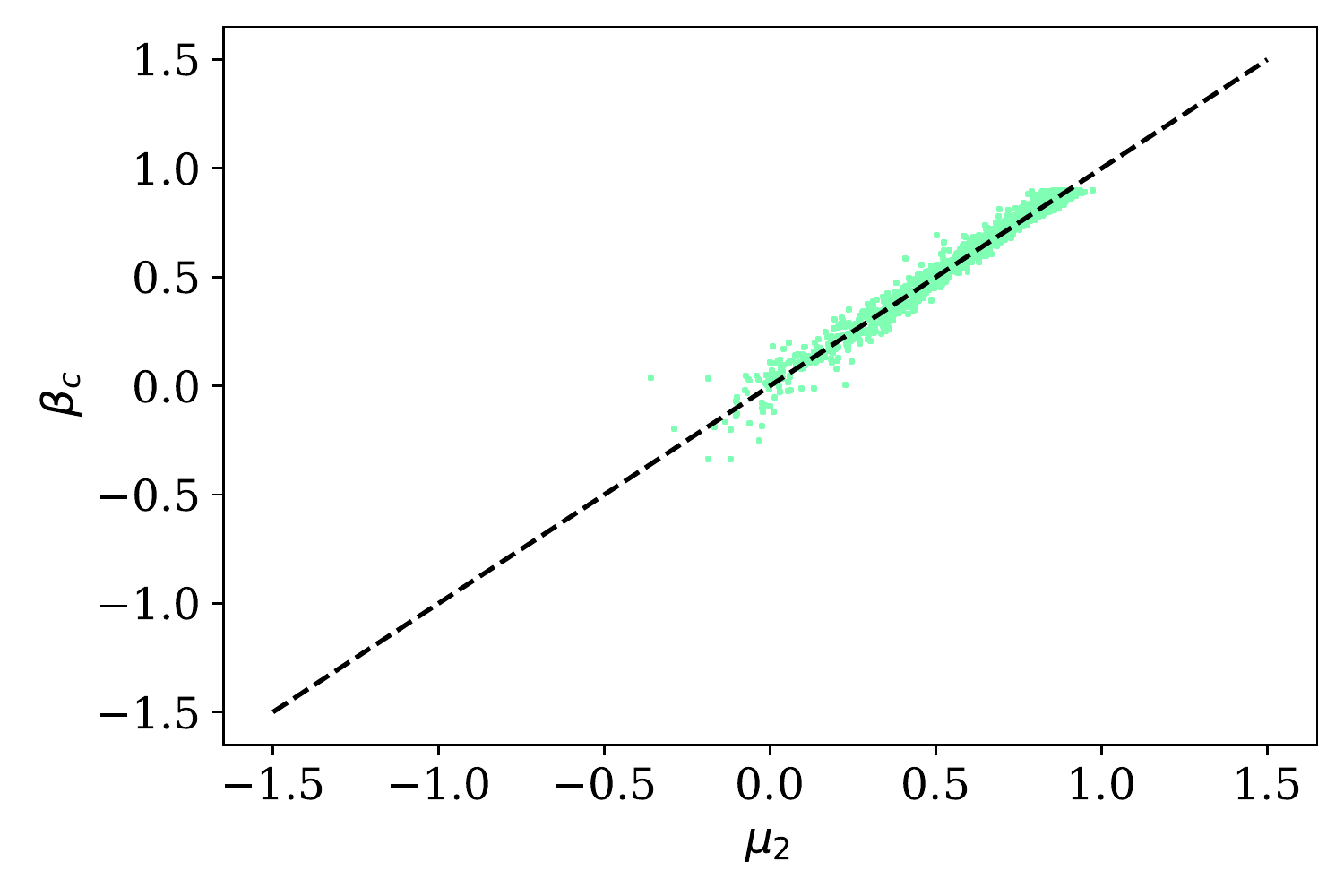}
        \caption{Cylinder}
        \label{fig:cylinder_train}
    \end{subfigure}
    \begin{subfigure}[b]{0.32\textwidth}
        \centering
        \includegraphics[width=\textwidth]{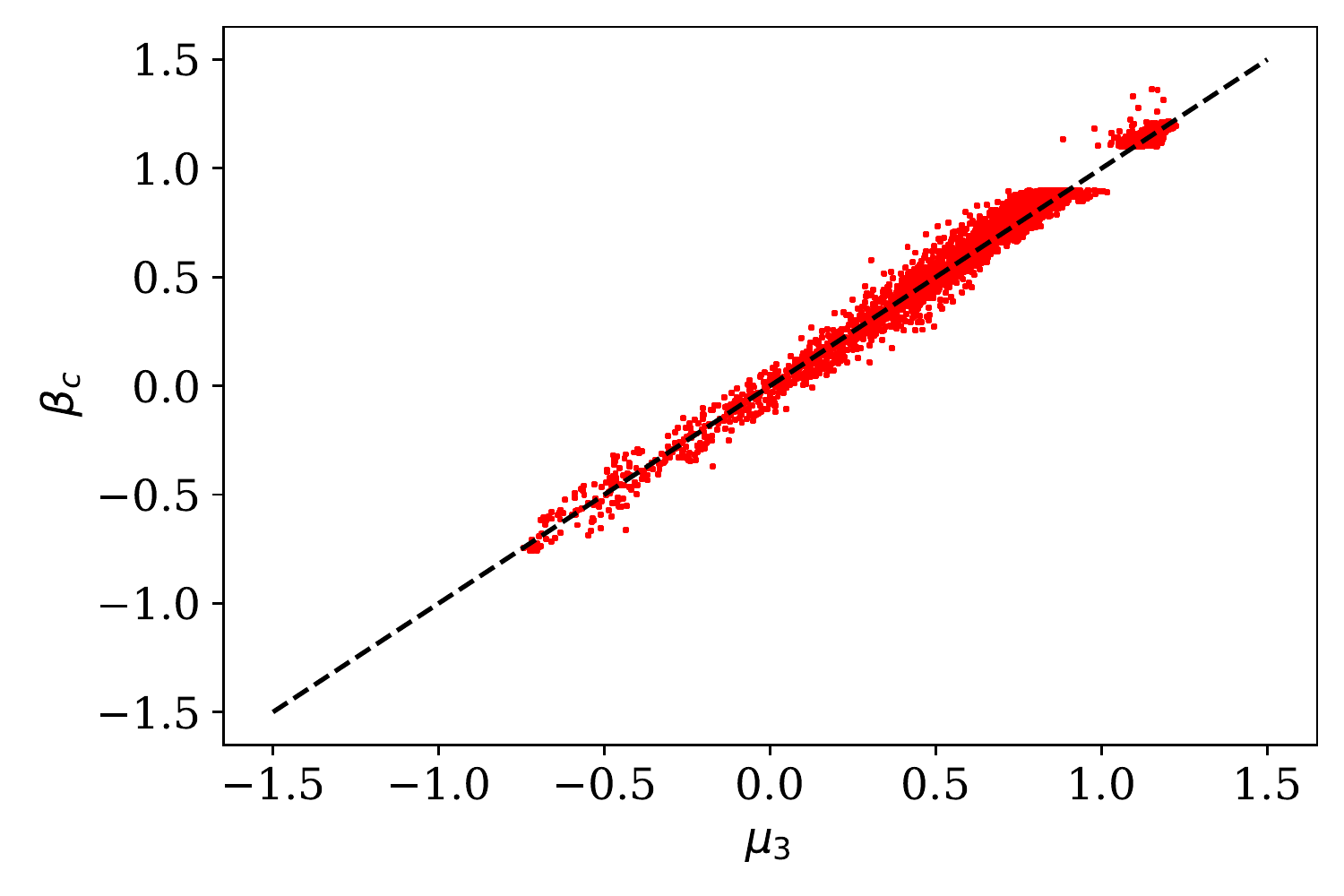}
        \caption{NASA hump}
        \label{fig:nasa_hump_train}
    \end{subfigure}

     \begin{subfigure}[b]{0.49\textwidth}
        \centering
        \includegraphics[width=\textwidth]{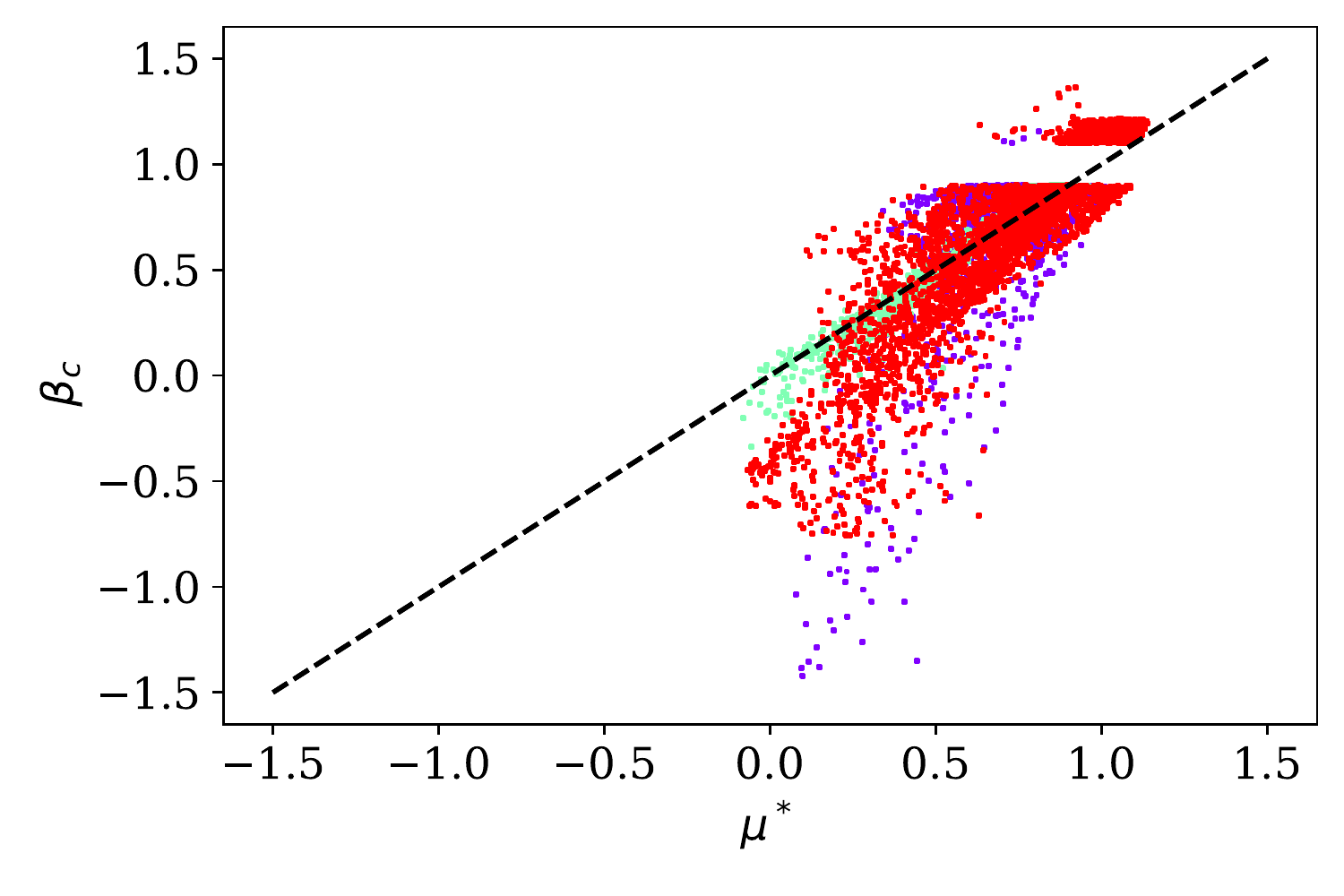}
        \caption{Entire ensemble}
        \label{fig:ensemble_train}       
    \end{subfigure}
    \begin{subfigure}[b]{0.49\textwidth}
        \centering
        \includegraphics[width=\textwidth]{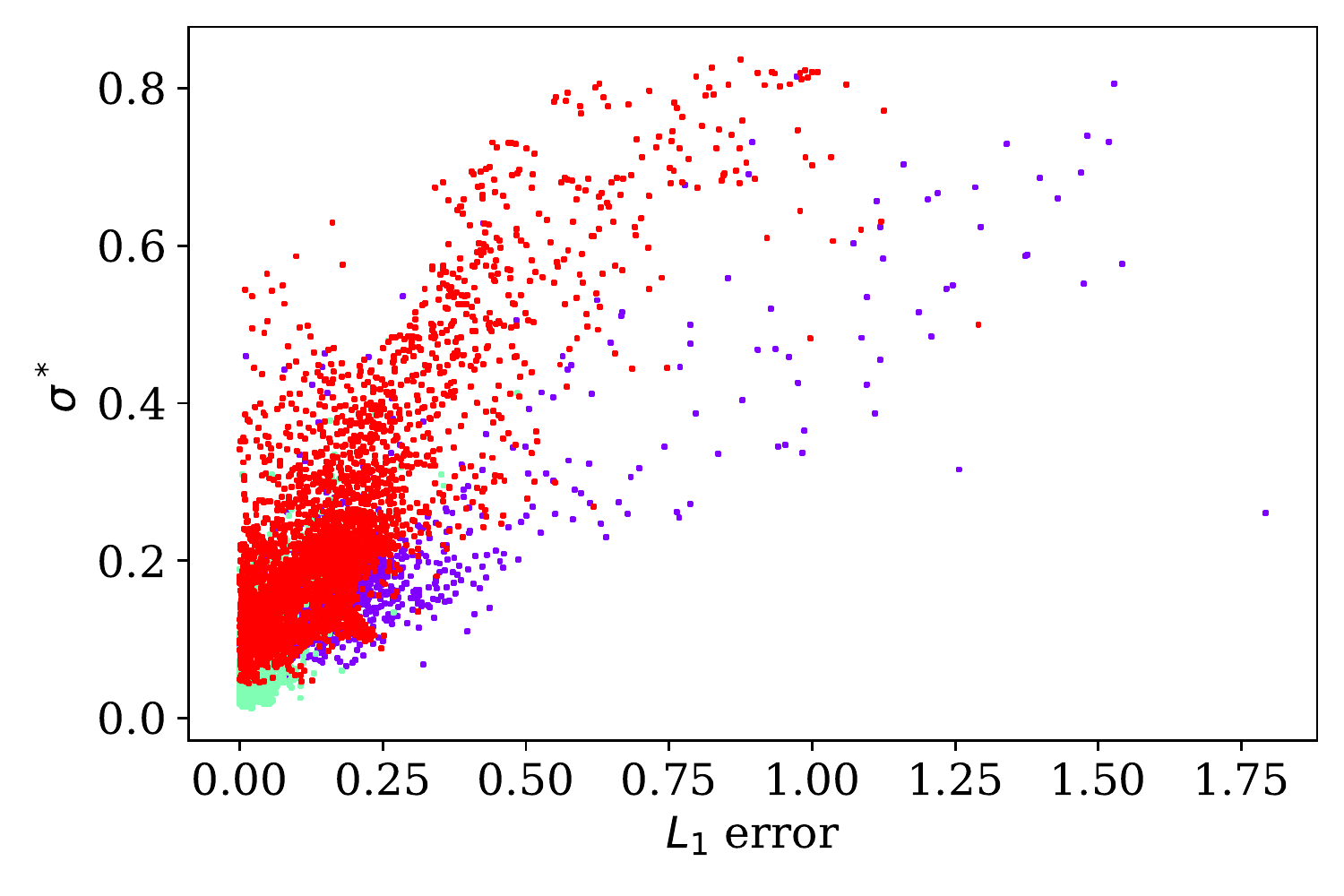}
        \caption{Ensemble $L_1$ error versus $\sigma^*$ }
        \label{fig:L1_sigma}       
    \end{subfigure}
    \caption{Results for the GPE ensemble trained on the inverted data: for the individual GPE submodels (a)-(c); the entire ensemble (d); and ensemble $L_1$ error against $\sigma^*$ (e).}
    \label{fig:gpe_train}
\end{figure}

Performing a field inversion for each of the three cases discussed above yielded the set of training data $D=[D_1, D_2, D_3]$, where each subset of the training data, $D_i$, collects the set of parameter points $(\boldsymbol{\alpha}^{(j)}, \beta_c^{(j)}),\, j=1,\dots, n_i$ harvested from the $i$\textsuperscript{th} case, where $n_i$ refers to the number of cells in which the correction is active. A GPE was trained on each subset of data, forming an ensemble of three GPEs. The results of the training are displayed in Figure \ref{fig:gpe_train}. Figures \ref{fig:bbfs_train}-\ref{fig:nasa_hump_train} show the results of the training for each of the GPEs in the ensemble, comparing the predictions of $\beta_c$ from the GPEs with the inverted results. Note there is a gap in the training data formed by removing parameter points in which there was not significant correction. Figure \ref{fig:ensemble_train} displays the results for the entire ensemble and Figure \ref{fig:L1_sigma} plots the ensemble error, as quantified by the $L_1$ error, against the predicted uncertainty, $\sigma^*$. 

\begin{figure}[h]
    \centering
     \begin{subfigure}[b]{0.49\textwidth}
        \centering
        \includegraphics[width=\textwidth]{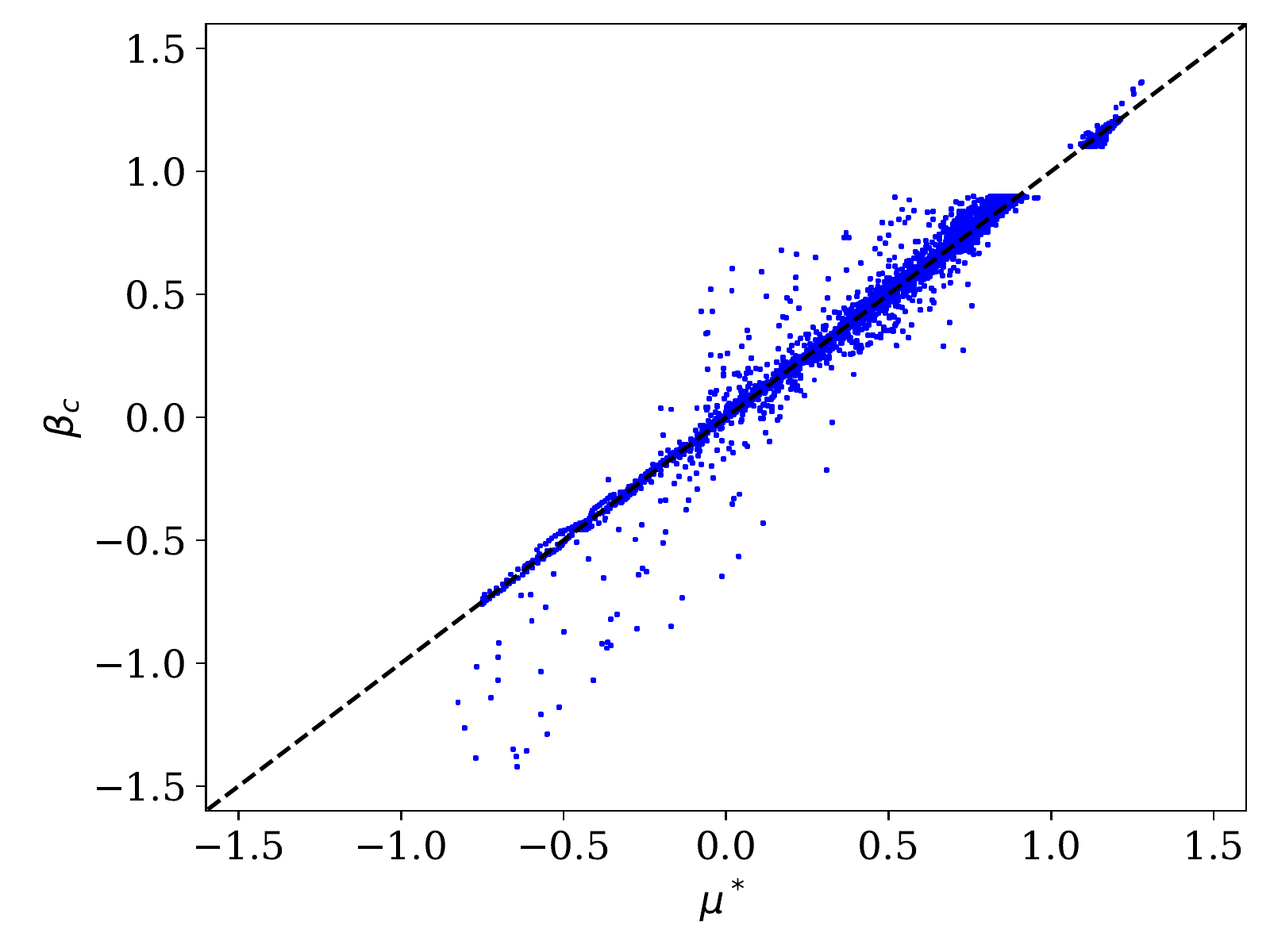}
        \caption{Deep ensembles}
        \label{fig:de_train}       
    \end{subfigure}
    \begin{subfigure}[b]{0.49\textwidth}
        \centering
        \includegraphics[width=\textwidth]{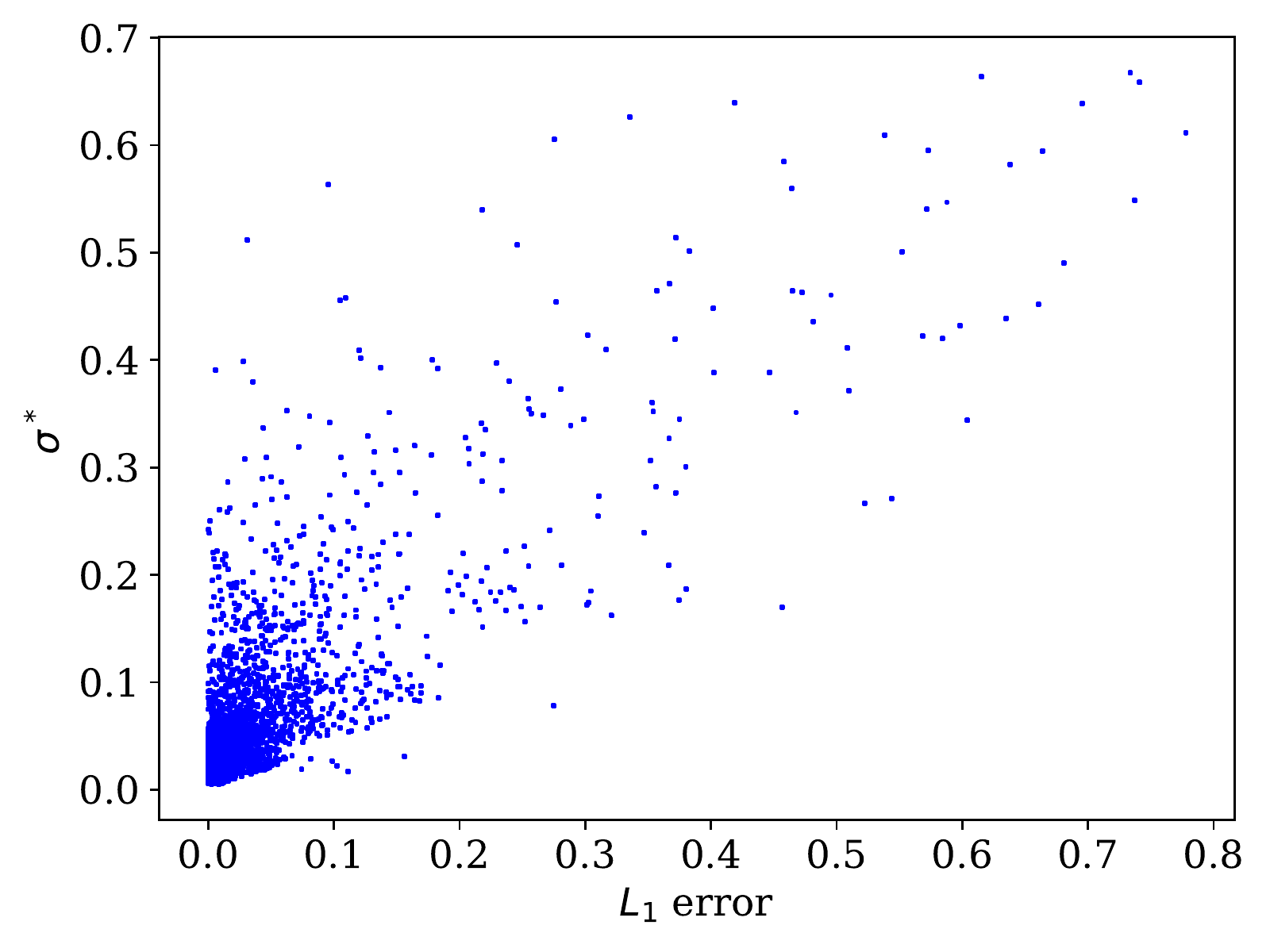}
        \caption{Deep ensembles $L_1$ error versus $\sigma^*$ }
        \label{fig:L1_sigma_de}       
    \end{subfigure}
    \caption{Results for the deep ensembles trained on the inverted data.}
    \label{fig:de_train}
\end{figure}

The training results for the deep ensembles is also shown in Figure \ref{fig:de_train}. Similar to the GPE, the $L_1$ error increases with the model prediction uncertainty. 

\section{Application to unseen test cases}
Two types of test cases were chosen to assess the efficacy of the probabilistic machine learning framework. The first type were cases where the ML models are expected to be active in certain regions of the flow, improving the results. For this type, the h42 bump of Matai and Durbin \cite{matai_durbin_2019} and the periodic hills case of Frohlich et al. \cite{frohlich_periodic_hills_2005} were tested. The second type of cases were those that do not include flow physics for which the corrections in the training data accounted for. Such cases include the zero pressure gradient flat plate and NACA0012 airfoil at various angles of attack, where the ML augmented turbulence model should revert to baseline.

\subsection{h42 bump}
A series of parametric bumps were simulated in LES by Matai and Durbin \cite{matai_durbin_2019} to study the systematic behaviour of the boundary layer and flow separation in increasing adverse pressure gradients. The flow was fully attached on the lowest bump while the highest bump had a separation bubble. Data from the paper was made available by McConkey \cite{mcconkey_curated_dataset}. The ML models were applied to the case with the highest bump (h42) and compared to field inverted results which were not used in training. The inversion was done to provide a reference to assess the ML model performance. 

The test case was taken from McConkey \cite{mcconkey_curated_dataset} with a grid of $412\times 175$ and a maximum $y^+$ of $0.85$ on the walls. At the maximum step height of $42\text{mm}$, the height based Reynolds number was $Re_h=27850$.

\subsubsection{Inversion results}
\begin{figure}[h]
     \centering
     \includegraphics[width=0.75\textwidth]{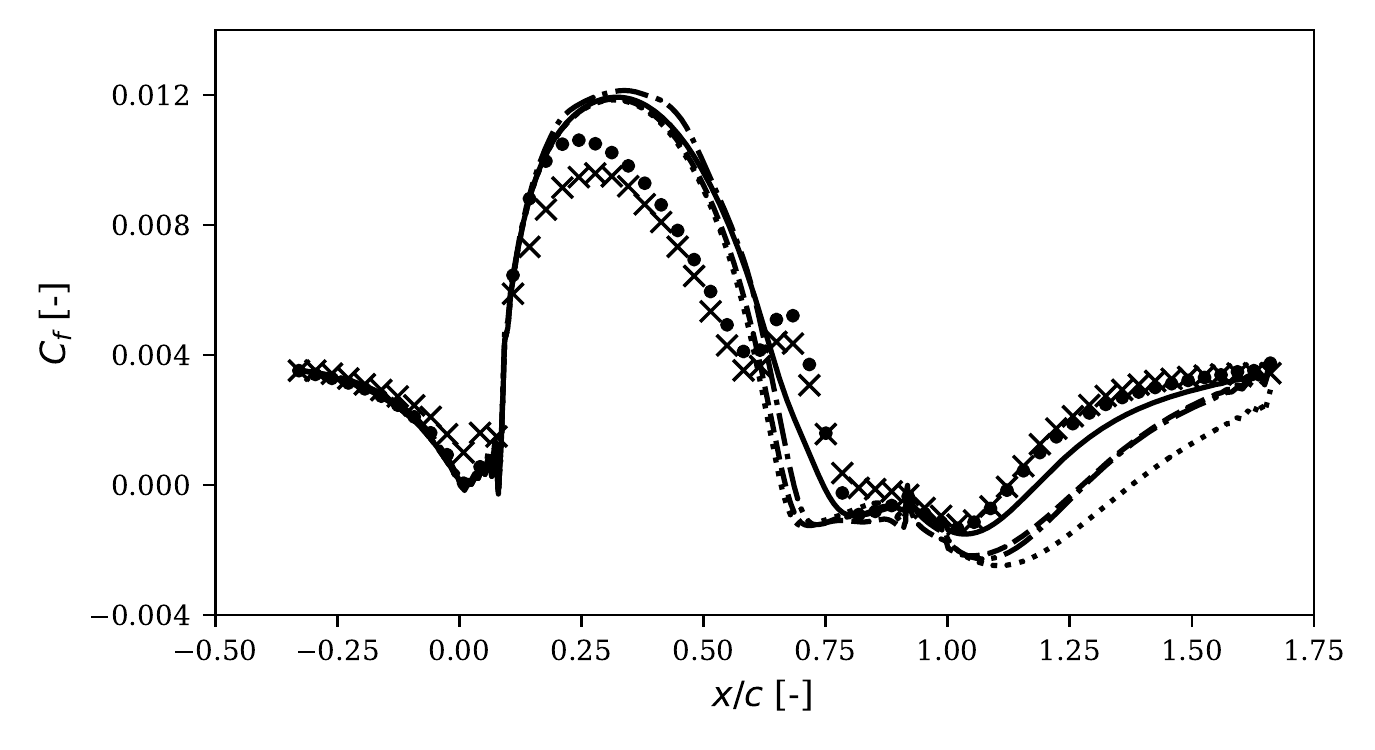}
     \caption{Skin friction on bump surface. Reference LES (\markercross), inverted (\markercircle), constrained inverted (\full), uncorrected (\dotted), GPEs (\dashed), DEs (\dotdash).}
     \label{fig:h42_cf}
\end{figure}

The full LES flow field was assimilated initially. It can be seen in Figure \ref{fig:h42_cf} that the separation region of the inverted results agree well with LES. The inverted results also showed a reduction in skin friction in the windward part of the bump along with the increase in the latter half contributed by an increase in turbulent kinetic energy near the wall before the flow separates.

\begin{figure}[h]
    \centering
    \begin{subfigure}[b]{0.32\textwidth}
        \centering
        \includegraphics[width=\textwidth]{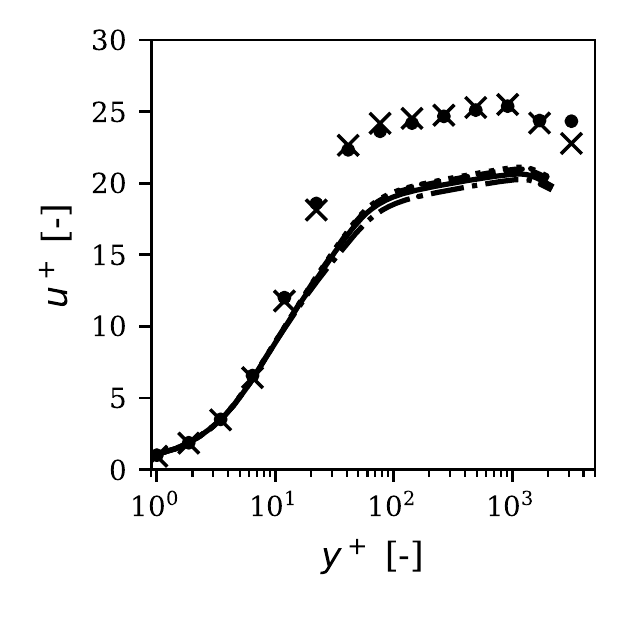}
        \caption{$x/c=0.50$.}
        \label{fig:bump_h42_all_uplus_0.50}
    \end{subfigure}
    \begin{subfigure}[b]{0.32\textwidth}
        \centering
        \includegraphics[width=\textwidth]{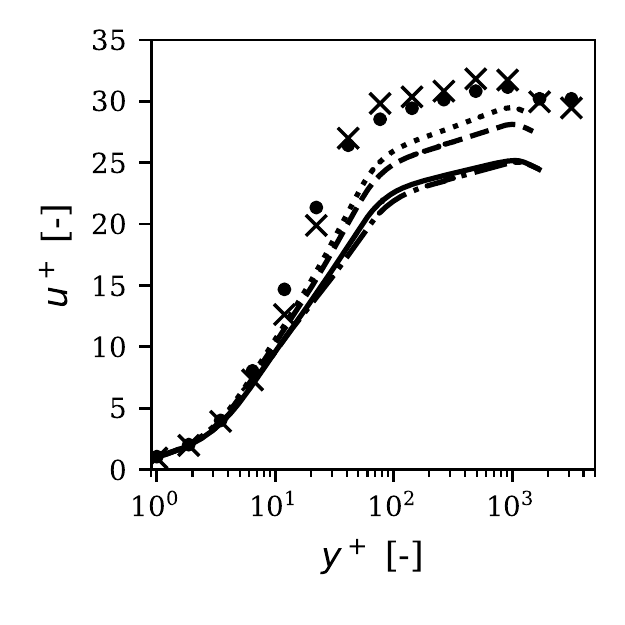}
        \caption{$x/c=0.60$.}
        \label{fig:bump_h42_all_uplus_0.60}
    \end{subfigure}
    
    \begin{subfigure}[b]{0.32\textwidth}
        \centering
        \includegraphics[width=\textwidth]{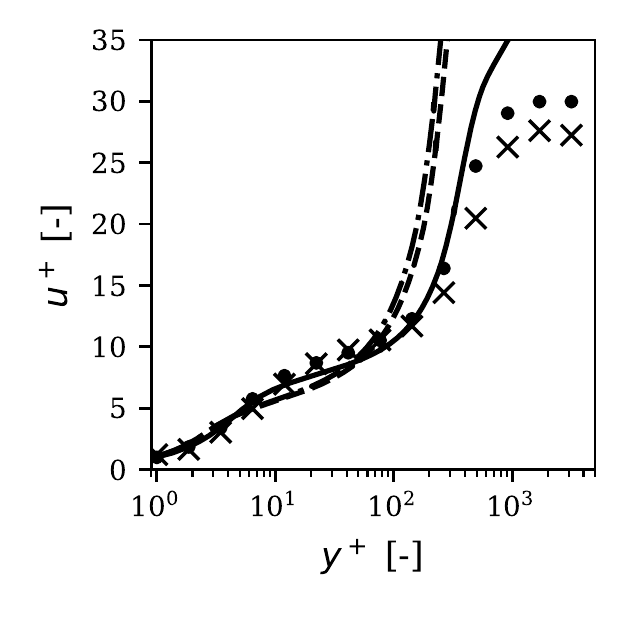}
        \caption{$x/c=1.33$.}
        \label{fig:bump_h42_all_uplus_1.33}
    \end{subfigure}
        \begin{subfigure}[b]{0.32\textwidth}
        \centering
        \includegraphics[width=\textwidth]{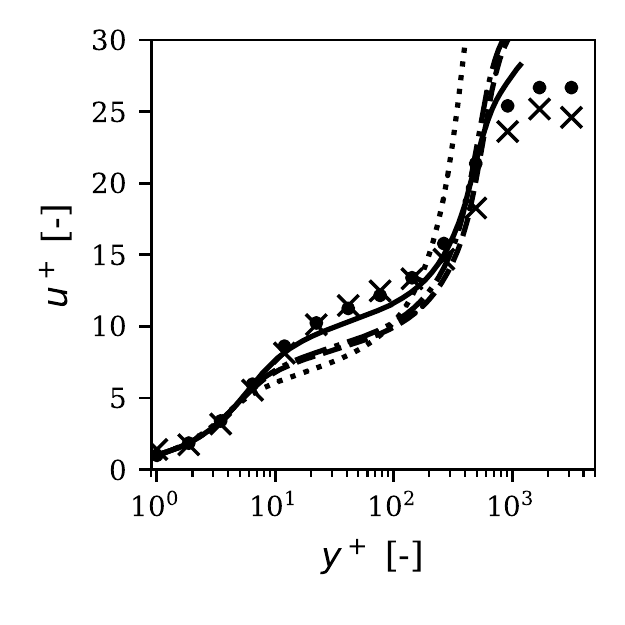}
        \caption{$x/c=1.50$.}
        \label{fig:bump_h42_all_uplus_1.50}
    \end{subfigure}
    \caption{Dimensionless velocity. See Figure \ref{fig:h42_cf} for legend.}
    \label{fig:h42_uplus}
\end{figure}

The non-dimensional velocity profile of the inverted results also compares well to that given in Figure \ref{fig:h42_uplus}. Similar to LES, the assimilated results show a thicker viscous sub layer at $x/c=0.5$ and $x/c=0.6$. There is also better flow recovery at $x/c=1.5$.

\begin{figure}[h]
    \centering
    \begin{subfigure}[b]{0.4\textwidth}
        \centering
        \includegraphics[width=\textwidth]{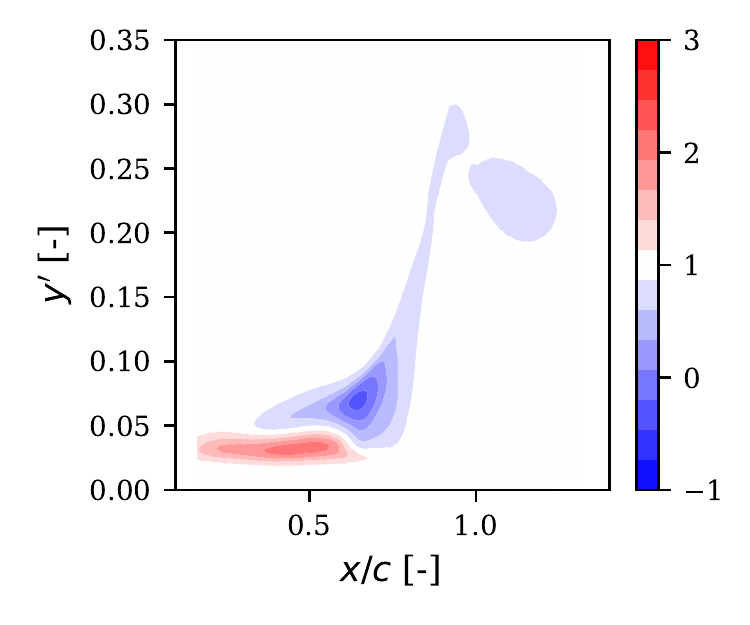}
        \caption{Inverted.}
        \label{fig:bump_h42_all_beta_invf}
    \end{subfigure}
    \begin{subfigure}[b]{0.4\textwidth}
        \centering
        \includegraphics[width=\textwidth]{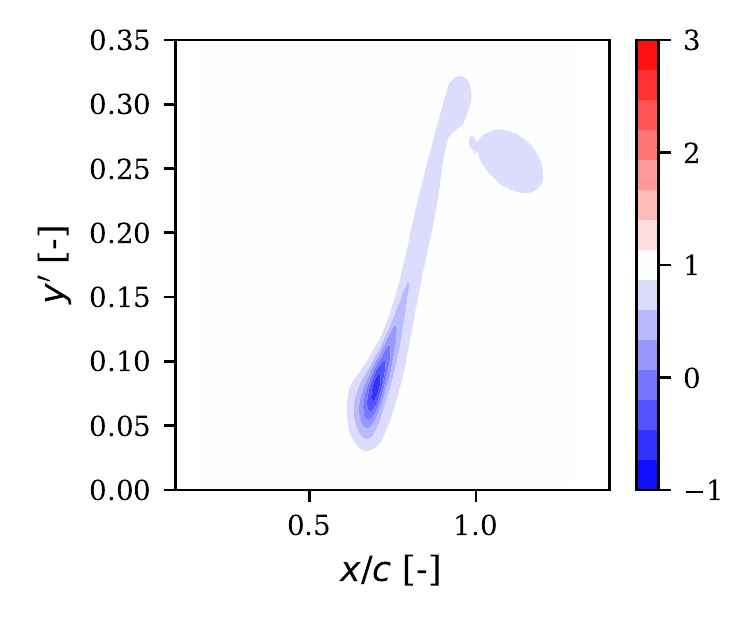}
        \caption{Constrained inverted.}
        \label{fig:bump_h42_all_beta_invp}
    \end{subfigure}
    
    \begin{subfigure}[b]{0.4\textwidth}
        \centering
        \includegraphics[width=\textwidth]{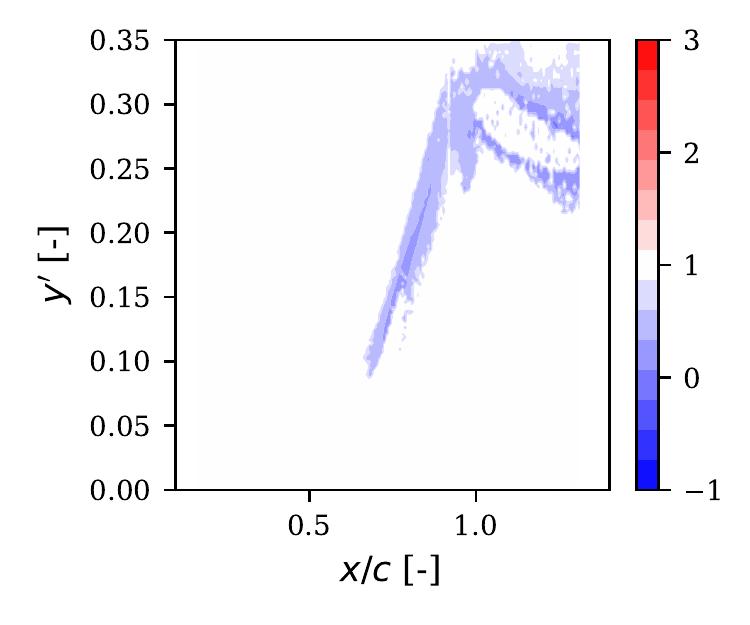}
        \caption{GPEs.}
        \label{fig:bump_h42_all_beta_gpe}
    \end{subfigure}
    \begin{subfigure}[b]{0.4\textwidth}
        \centering
        \includegraphics[width=\textwidth]{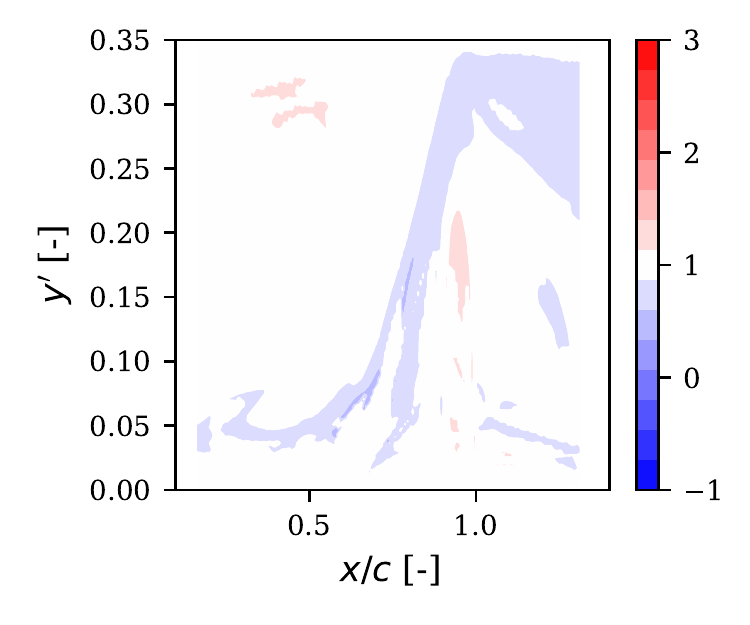}
        \caption{DEs.}
        \label{bump_h42_all_beta_de}
    \end{subfigure}
    \caption{Inverted and predicted $\beta_c$ fields. Scaled $y$ coordinates used where $y^{\prime}=\sqrt{(y-y_{wall})/c}$.}
    \label{fig:h42_beta}
\end{figure}

The RANS results over-predicted shear stress around $x/c<0.6$ before under-predicting past that point. The inverted correction field in Figure \ref{fig:h42_beta} show an increased $\beta_c$ in the near wall around $x<0.6$ to increase the specific dissipation rate hence reducing $k$ and $\nu_t$. This leads to a reduction in shear stress. It was noted in \cite{matai_durbin_2019} that the flow does not relaminarise. The uncorrected RANS shear stress becomes under-predicted after $x>0.6$. There is also a region of $\beta_c<1$ off the wall surface causing an increase in $k$ which reduces the separation bubble.

The ML models are not expected to predict the increased $\beta_c$ in the forward part of the bump since the training data is made up largely of $\beta_c<1$ in the separated region. A constrained field inversion was performed where $\beta_c$ was only allowed to be active from $x/c>0.6$ and a smaller region of LES data in the separated region was assimilated (i.e. $0.8<x/C<1.5$ and $0<y/C<0.16$). This is more representative of an inversion procedure used to generate the training data which can then be used to assess the ML models. The constrained inversion only had a region of reduced $\omega$ production in the separated region (Figure \ref{fig:h42_beta}) without the drop in $k$ in the near wall. %\textcolor{red}{@Joel say something about the velocity profile going the other way at $x/c=0.6$, something about how it correlates with the overprediction in $Cf$ in that region?}.

\subsubsection{Variance tolerance sweep}

\begin{figure}[h]
    \centering
    \begin{subfigure}[b]{0.475\textwidth}
        \centering
        \includegraphics[width=\textwidth]{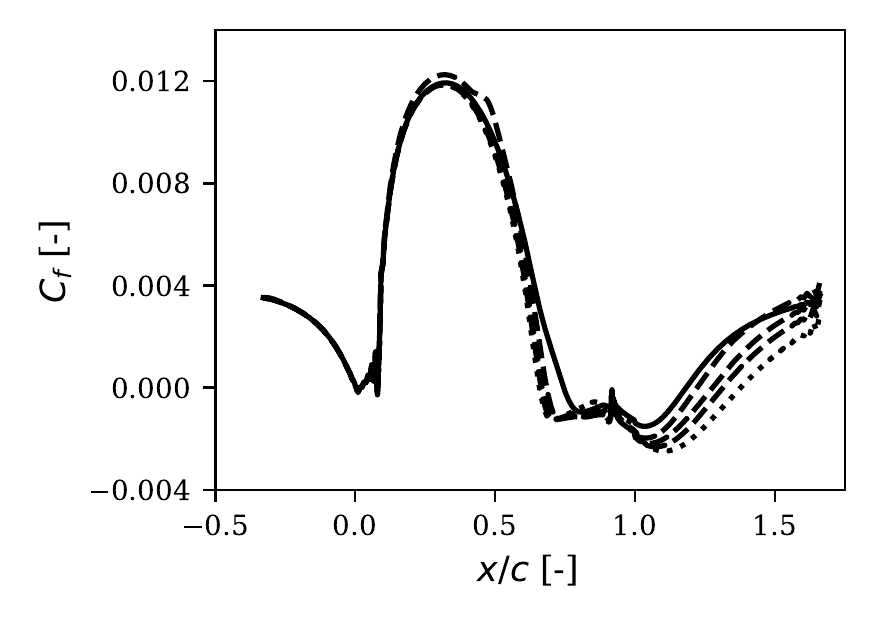}
        \caption{GPEs results for $\bar{\sigma}=[0.15,0.2,0.25]$.}
        \label{fig:bump_h42_gpe_sigma_sweep}
    \end{subfigure}
    \begin{subfigure}[b]{0.475\textwidth}
        \centering
        \includegraphics[width=\textwidth]{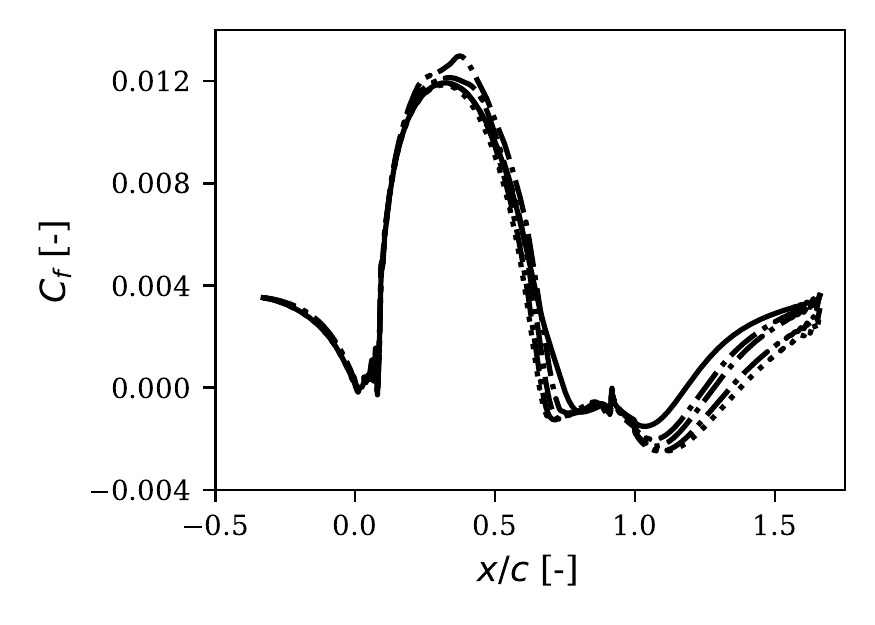}
        \caption{DEs results for $\bar{\sigma}=[0.05,0.1,0.15]$.}
        \label{fig:bump_h42_de_sigma_sweep}
    \end{subfigure}
    \caption{Skin friction coefficient. Constrained inverted (\full), uncorrected (\dotted), GPEs (\dashed), DEs (\dotdash). Results from lower values of $\bar{\sigma}$ are closer to the uncorrected results. Increasing $\bar{\sigma}$ brings profile closer to constrained inverted results but also adds overshoot after a certain point.}
    \label{fig:h42_sigma_sweep}
\end{figure}

A sweep of $\bar{\sigma}$ was done on both models to assess the effects of adjusting the acceptance criterion of \eqref{eq:crit}. Figure \ref{fig:bump_h42_gpe_sigma_sweep} shows how the skin friction profile changes from the uncorrected profile towards the constrained inverted results for the GPEs with increasing $\bar{\sigma}$. At a low $\bar{\sigma}$ of $0.15$, there is a slight improvement on the reattachment length while having no upstream effects. The $C_f$ profile remains bounded between the constrained inverted and uncorrected results until $\bar{\sigma}=0.25$, where the reattachment point matches that of the constrained inverted results, but $C_f$ overshoots the bounds starting at $x/c=0.25$ on the windward face. A tolerance of $0.2$ was hence chosen for the GPEs for all of the test cases.

Figure \ref{fig:bump_h42_de_sigma_sweep} shows a similar behaviour with the DEs. However, it overshoots the $C_f$ much earlier than GPEs with only the $\bar{\sigma}=0.05$ results remaining bounded. At that tolerance there was little improvements to the reattachment length. To reach the same amount of improvements to the reattachment length as the GPEs at its prescribed tolerance, a $\bar{\sigma}$ value of $0.1$ was assigned to the DEs allowing some overshoot over the hump. These values were kept constant for all test cases in this work.

\subsubsection{ML predictions}

\begin{figure}
    \centering
    \begin{subfigure}[b]{0.32\textwidth}
        \centering
        \includegraphics[width=\textwidth]{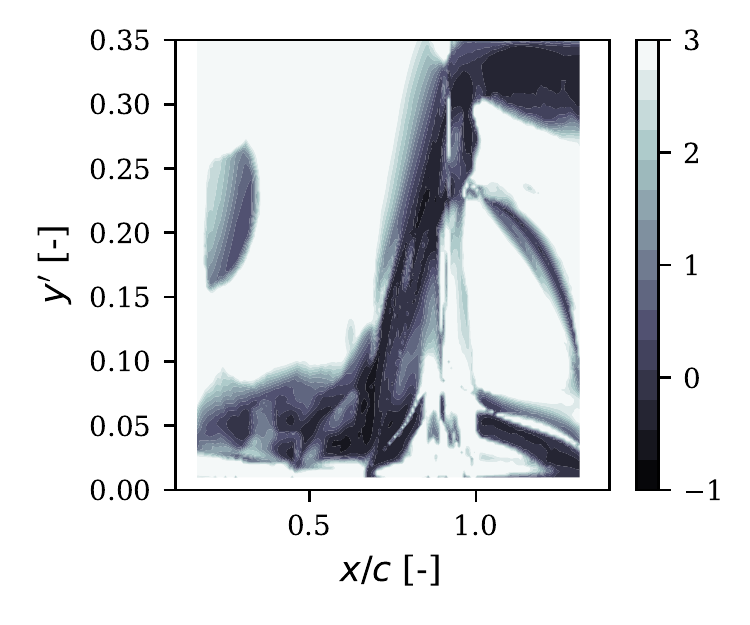}
        \caption{Local outlier factor.}
        \label{fig:bump_h42_all_lof}
    \end{subfigure}
    
    \begin{subfigure}[b]{0.32\textwidth}
        \centering
        \includegraphics[width=\textwidth]{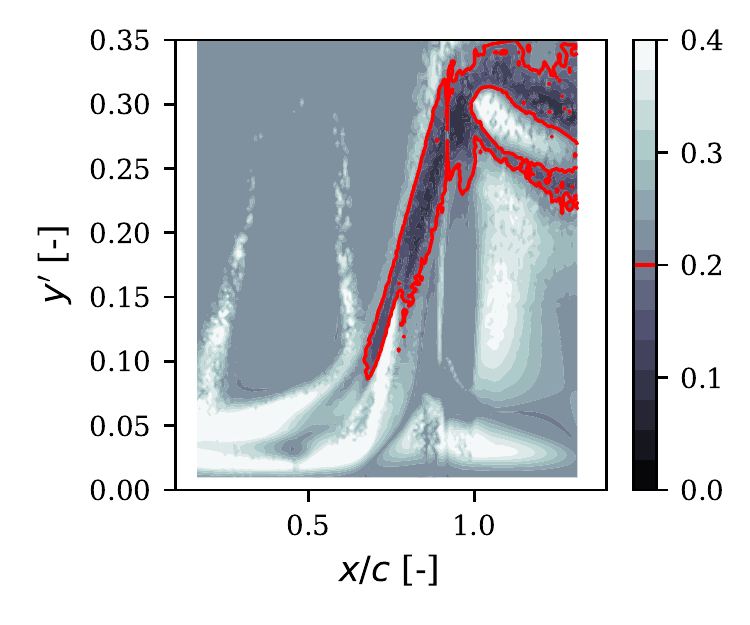}
        \caption{$\sigma^*$ from GPEs.}
        \label{fig:bump_h42_all_gpe_sqrt_var}
    \end{subfigure}
    \begin{subfigure}[b]{0.32\textwidth}
        \centering
        \includegraphics[width=\textwidth]{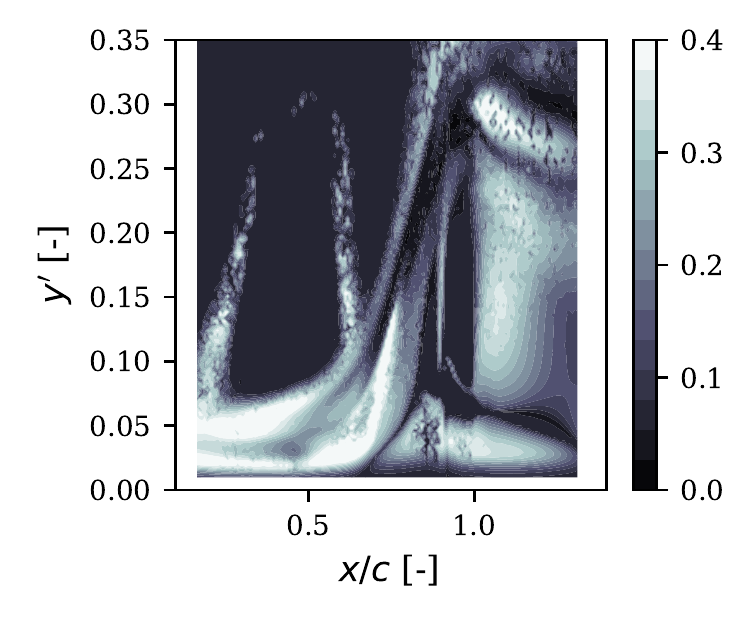}
        \caption{$\sigma^*_{\mu}$ GPEs}
        \label{fig:bump_h42_all_gpe_sqrt_var_mu}
    \end{subfigure}
    \begin{subfigure}[b]{0.32\textwidth}
        \centering
        \includegraphics[width=\textwidth]{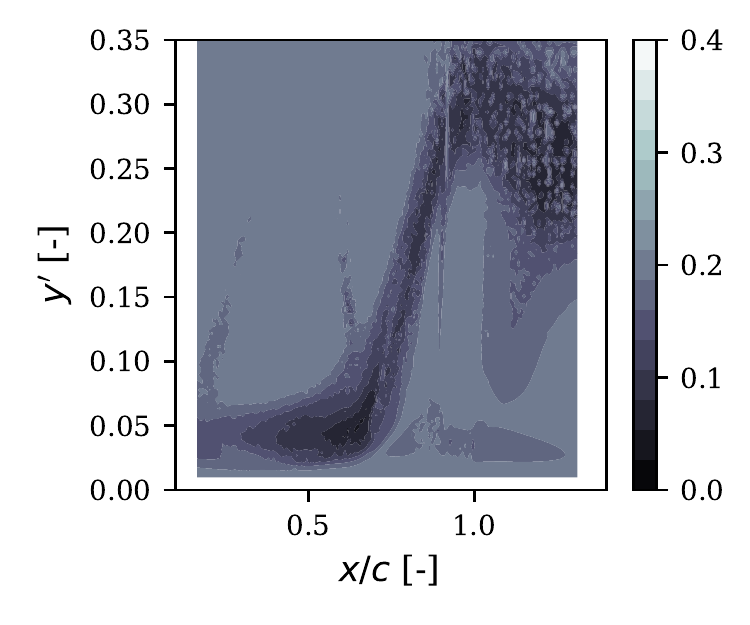}
        \caption{$\sigma^*_{\sigma}$ GPEs}
        \label{fig:bump_h42_all_gpe_sqrt_var_sigma}
    \end{subfigure}
    
    \begin{subfigure}[b]{0.32\textwidth}
        \centering
        \includegraphics[width=\textwidth]{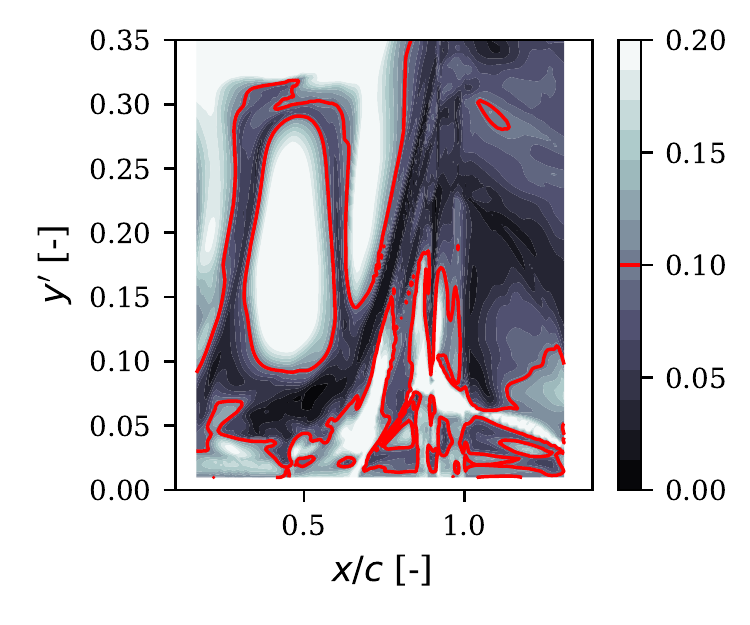}
        \caption{$\sigma^*$ from DEs.}
        \label{fig:bump_h42_all_de_sqrt_var}
    \end{subfigure}
        \begin{subfigure}[b]{0.32\textwidth}
        \centering
        \includegraphics[width=\textwidth]{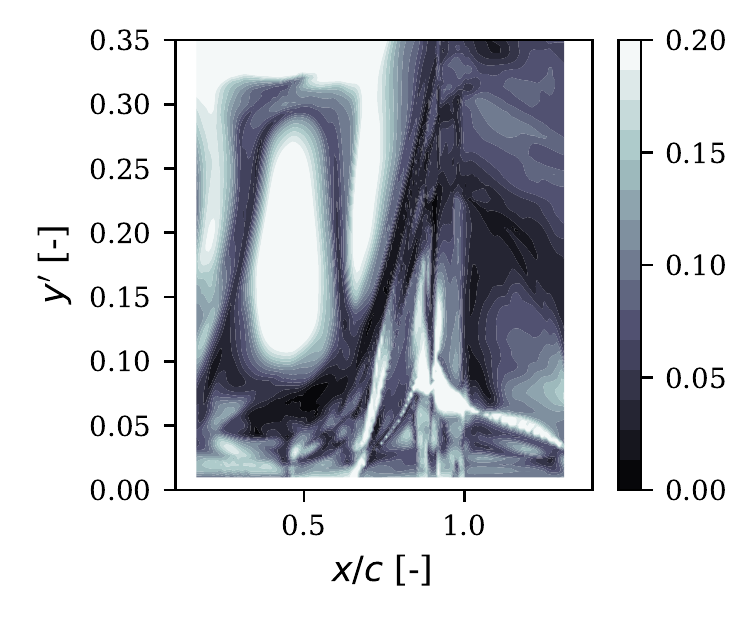}
        \caption{$\sigma^*_{\mu}$ DEs}
        \label{fig:bump_h42_all_de_sqrt_var_mu}
    \end{subfigure}
        \begin{subfigure}[b]{0.32\textwidth}
        \centering
        \includegraphics[width=\textwidth]{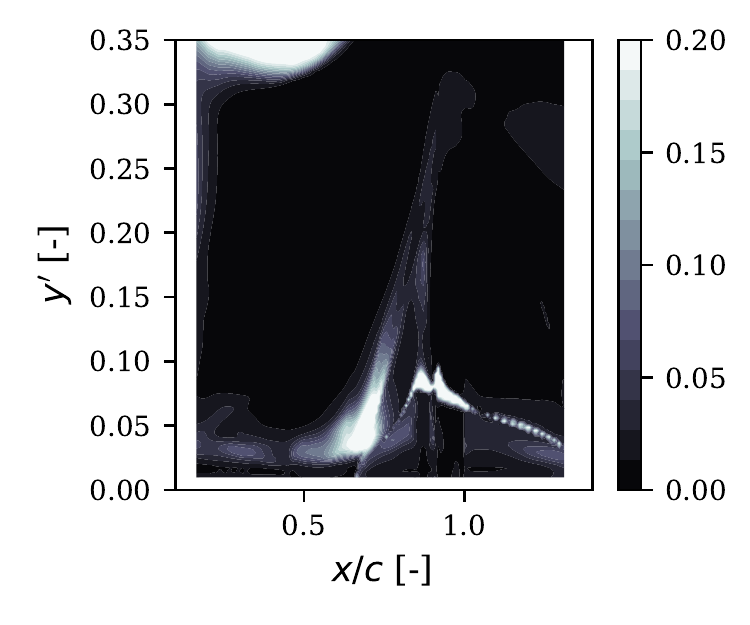}
        \caption{$\sigma^*_{\sigma}$ DEs}
        \label{fig:bump_h42_all_de_sqrt_var_sigma}
    \end{subfigure}
    \caption{Extrapolation detection and variances from ML models. (\subref{fig:bump_h42_all_gpe_sqrt_var_mu}) $\sigma^*_{\mu}$ from GPEs representing the spread of $\beta_c$ across the different inverted cases. (\subref{fig:bump_h42_all_gpe_sqrt_var_sigma}) $\sigma^*_{\sigma}$ from GPEs representing the average variance within each inverted case. Low values indicate that test feature is close to training feature while large values show where model is extrapolating. (\subref{fig:bump_h42_all_de_sqrt_var_mu}) $\sigma^*_{\mu}$ from DEs representing the spread of means of the submodels, indicating how much the model is extrapolating. (\subref{fig:bump_h42_all_de_sqrt_var_sigma}) $\sigma^*_{\sigma}$ from DEs representing the spread of $\beta_c$ in the training data collected across all inverted cases.}
    \label{fig:h42_variance}
\end{figure}

\begin{figure}[h]
    \centering
    \begin{subfigure}[b]{0.475\textwidth}
        \centering
        \includegraphics[width=\textwidth]{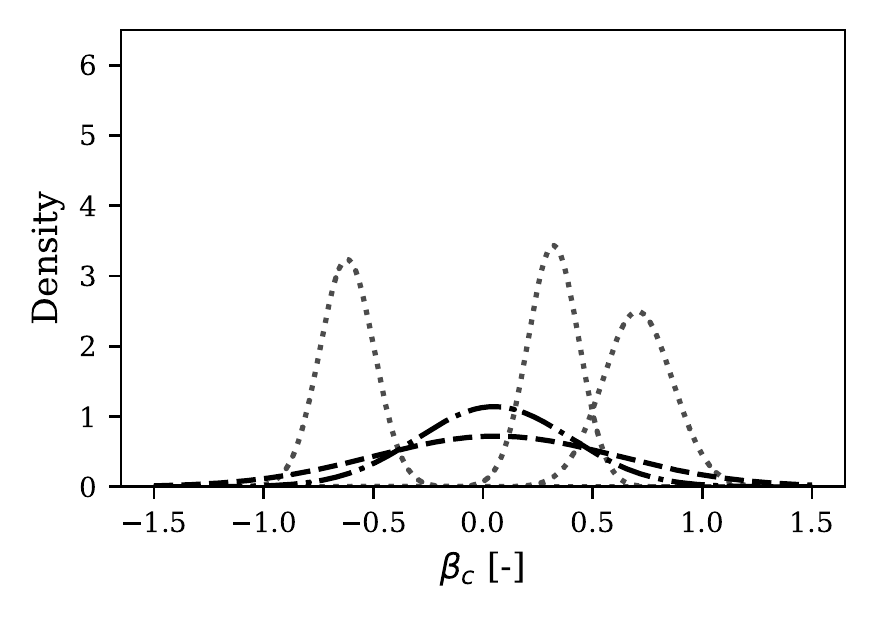}
        \caption{$x/c=0.69, y^{\prime}=0.06$.}
        \label{fig:bump_h42_all_high_var_1}
    \end{subfigure}
    \begin{subfigure}[b]{0.475\textwidth}
        \centering
        \includegraphics[width=\textwidth]{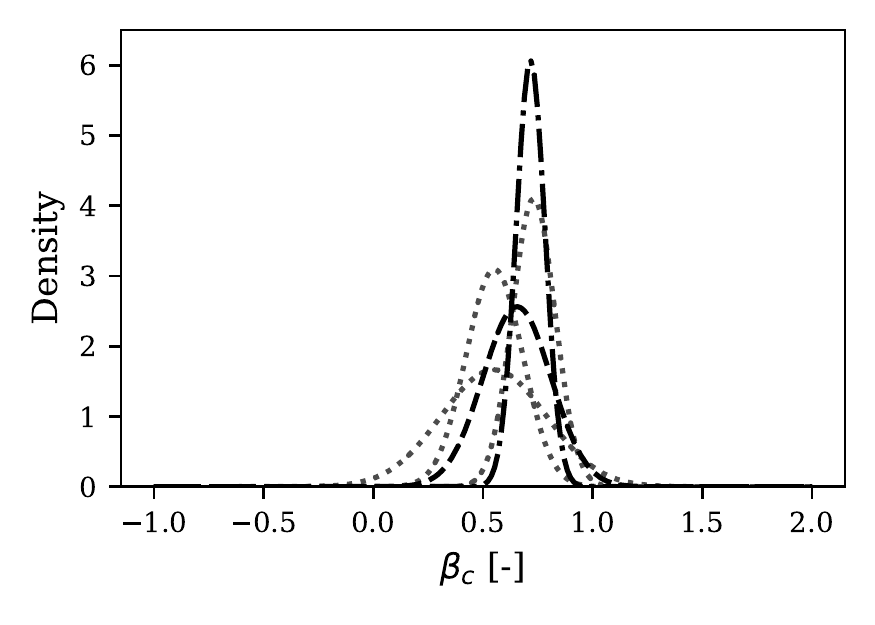}
        \caption{$x/c=0.90, y^{\prime}=0.23$.}
        \label{fig:bump_h42_all_low_var_0}
    \end{subfigure}
    \caption{Gaussian outputs. GPEs submodel outputs (\dotted), GPEs ensemble (\dashed), DEs (\dotdash).}
    \label{fig:h42_gaussian}
\end{figure}

Both the GPEs and DEs predicted a region of reduced $\beta_c$ in the separated region, improving on the deficiency of the turbulence model as described in Section \ref{sec:methodology}. Both models were able to improve the reattachment and flow recovery after separation. Predictions from the GPEs were active further off the wall than DEs, with the $C_f$ overlapping that of the uncorrected results until $x/c=1.1$ after which it improves on it. Unlike the constrained inverted results, both ML models did not predict the delayed separation. The $C_f$ profile of the GPEs model was bounded by the uncorrected and constrained inversion results. Due to the greater activation of the DEs in the near wall, it diverged from the bounds around $0.25<x/c<0.6$. 

Figure \ref{fig:h42_variance} shows the variance predicted from the models. The red contours show regions where the ML model was below the prescribed threshold and thus active. The variances were compared to the results from a novelty detection analysis done through local outlier factor (LOF), with lower scores indicating points closer to the training data set \cite{Breunig_LOF}. Both the GPEs and DEs were active in the regions of low LOF away from the wall where $y^{\prime}>0.1$. Variance from the GPEs correlate well with LOF in that region, with low variance in the band of low LOF around $x/c>0.75$. The band of low LOF corresponds to the shear layer above the separation bubble. 

In the region $y^{\prime}<0.1$ where LOF is low, the GPEs did not predict a low variance. One particular point of interest is $x/c=0.69, y^{\prime}=0.06$, where the LOF is low, but variance from both GPEs and DEs were both high. The variance was broken down into $\sigma^*_{\sigma}$ and $\sigma^*_{\mu}$ in Figure \ref{fig:h42_variance}. GPEs had low $\sigma^*_{\sigma}$ which indicates that high confidence can be assigned to the predictions from the sub models trained on individual cases. However, $\sigma^*_{\mu}$ was high which indicates that the predicted $\beta_c$ values from each of the training cases were conflicting with each other. DEs at that point has a high $\sigma^*_{\sigma}$, which indicates that the training $\beta_c$ values were highly scattered at that point, further suggesting that there is a one-to-many mapping of $\beta_c$ within the different training cases at that point. This is confirmed in Figure \ref{fig:bump_h42_all_high_var_1}, showing the highly confident yet scattered predictions from each sub model of the GPEs and the subsequent low confidence from both the ensembles. Figure \ref{fig:bump_h42_all_low_var_0} shows the distributions from another point with high confidence where the predicted $\beta_c$ from the GPEs sub model are clustered together leading to low variance from both ensembles. 

Figure \ref{fig:bump_h42_all_gpe_sqrt_var_sigma} shows the asymptotic value of $\sigma^*_\sigma$ based on the prior lengthscales of the individual GPEs (i.e. $\sigma_{\sigma}^*$ approaches a maximum constant value for test points far away from the training data set). Any further increases to $\sigma^*$ at that point stems from the one-to-many mapping of $\beta_c$. As such, the asymptotic $\sigma_{\sigma}^*$ value provides an upper bound for the tolerance $\bar{\sigma}$ and is a good starting point to tune $\bar{\sigma}$.

Figure \ref{fig:bump_h42_all_de_sqrt_var_sigma} shows that in regions where data was present, DEs was able to learn the spread of $\beta_c$ in the training data. However, in areas of high LOF where it's extrapolating, $\sigma^*_{\sigma}$ drops to a low value. The total variance in these areas is largely dependent on $\sigma^*_{\mu}$, which in turn depends on the randomisation of the initial weights in the model.

\FloatBarrier
\subsection{Periodic hills}
The periodic hills case of \cite{frohlich_periodic_hills_2005} was chosen as it contains a region of massive separation off a curved surface similar to the training cases. Improvements from the ML models are thus expected. Similar to the h42 bump case, a field inversion was performed to create a set of reference results (which was not used in training) for assessment of the ML models. The simulation was set up with periodic boundary conditions around the crest of the hill with $Re_h=10595$ based on the height of the hill and bulk velocity at the crest. A structured block mesh of $300\times 180$ was used with a maximum $y^+$ of 0.6 at the walls. An unconstrained inversion was done to create the reference results with data from the LES of \cite{frohlich_periodic_hills_2005} assimilated in the region $0.46<x/h<4.6$ and $0<y/h<1.78$.

%\begin{figure}[h]
 %   \centering
 %   \begin{subfigure}[b]{0.49\textwidth}
%        \centering
%        \includegraphics[width=\textwidth]{images/GPE/betas_ensemble_phills_sigma.pdf}
%        \caption{}
%        \label{fig:ph_gpe1}
%    \end{subfigure}
%    \begin{subfigure}[b]{0.49\textwidth}
%        \centering
%        \includegraphics[width=\textwidth]{images/GPE/betas_ensemble_phills_L1.pdf}
%        \caption{}
%        \label{fig:ph_gpe2}
%    \end{subfigure}
%    \caption{}
%    \label{fig:ph_ensemble}
%\end{figure}

\begin{figure}[h]
    \centering
    \begin{subfigure}[b]{0.32\textwidth}
        \centering
        \includegraphics[width=\textwidth]{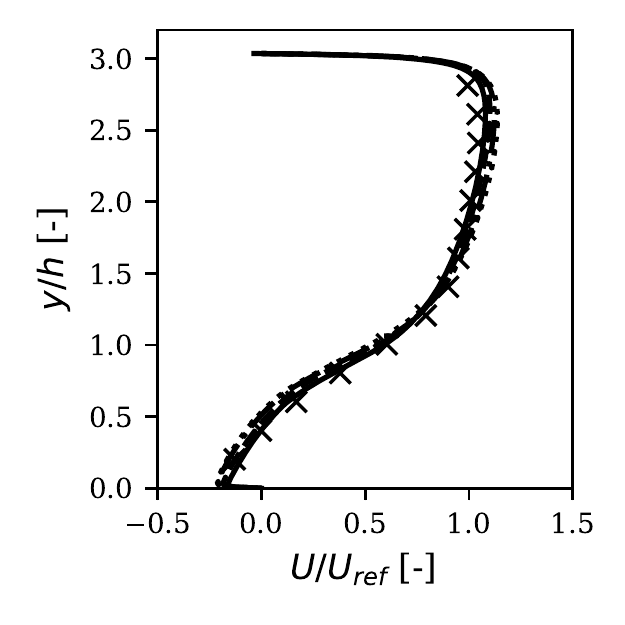}
        \caption{$x/h=3$}
        \label{fig:ph_vel_3}
    \end{subfigure}
    \begin{subfigure}[b]{0.32\textwidth}
        \centering
        \includegraphics[width=\textwidth]{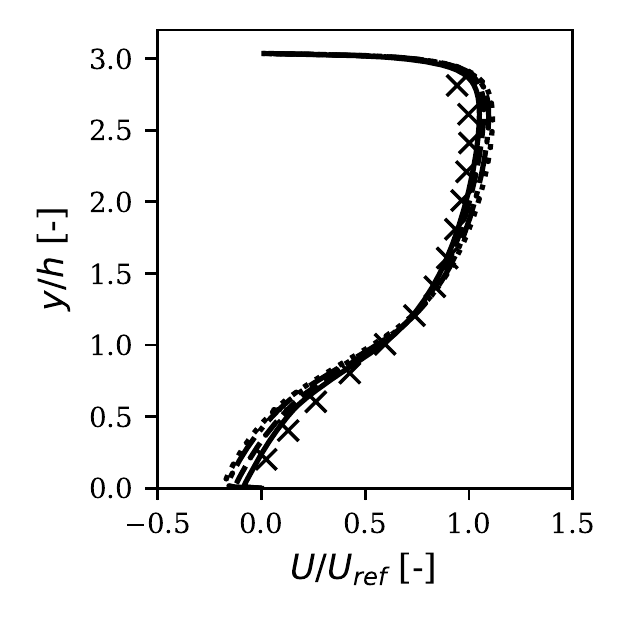}
        \caption{$x/h=4$}
        \label{fig:ph_vel_4}
    \end{subfigure}
    \begin{subfigure}[b]{0.32\textwidth}
        \centering
        \includegraphics[width=\textwidth]{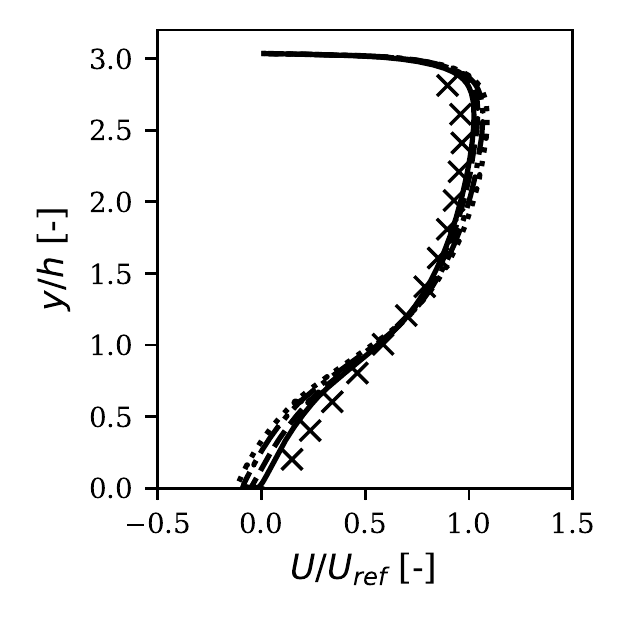}
        \caption{$x/h=5$}
        \label{fig:ph_vel_5}
    \end{subfigure}
    
    \begin{subfigure}[b]{0.32\textwidth}
        \centering
        \includegraphics[width=\textwidth]{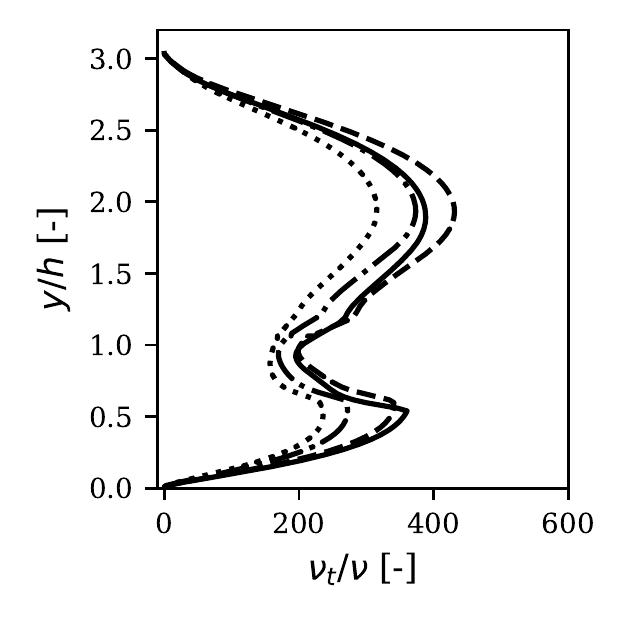}
        \caption{$x/h=3$}
        \label{fig:ph_nut_3}
    \end{subfigure}
    \begin{subfigure}[b]{0.32\textwidth}
        \centering
        \includegraphics[width=\textwidth]{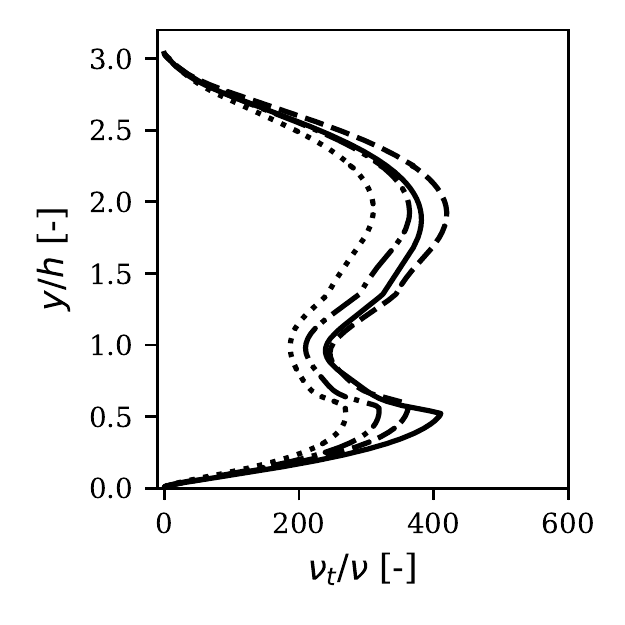}
        \caption{$x/h=4$}
        \label{fig:ph_nut_4}
    \end{subfigure}
    \begin{subfigure}[b]{0.32\textwidth}
        \centering
        \includegraphics[width=\textwidth]{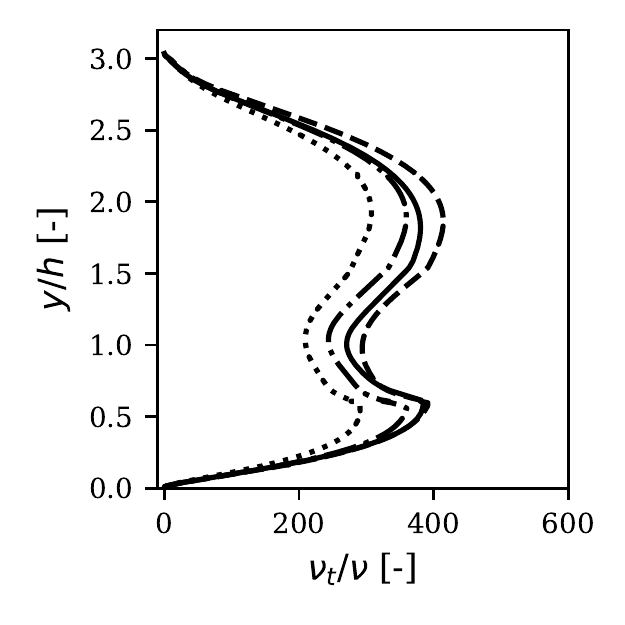}
        \caption{$x/h=5$}
        \label{fig:ph_nut_5}
    \end{subfigure}
    \caption{Periodic hills velocity profile and $\nu_t$ ratio. LES (\markercross), inverted (\full), uncorrected (\dotted), GPEs (\dashed), DEs (\dotdash). $\nu_t$ ratio compared against inverted results and not from LES.}
    \label{fig:ph_vel_nut}
\end{figure}

\begin{figure}[h]
    \centering
    \begin{subfigure}[b]{0.475\textwidth}
        \centering
        \includegraphics[width=\textwidth]{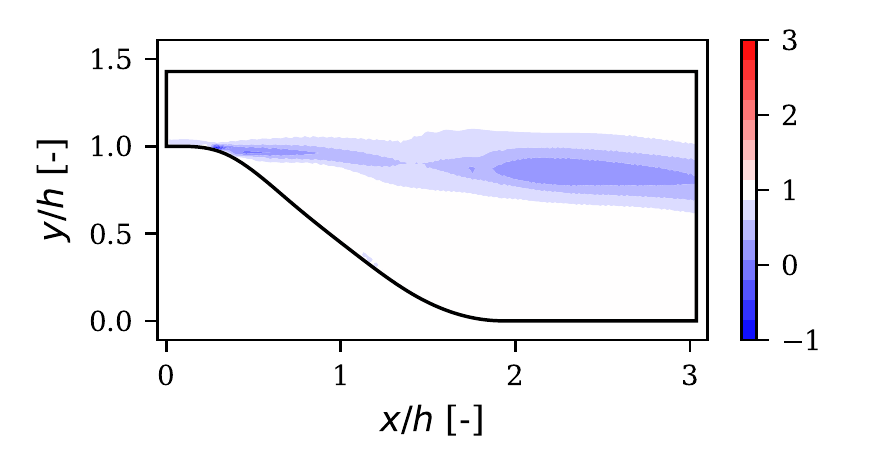}
        \caption{Inverted $\beta_c$ field.}
        \label{fig:ph_beta_inv}
    \end{subfigure}
    
    \begin{subfigure}[b]{0.475\textwidth}
        \centering
        \includegraphics[width=\textwidth]{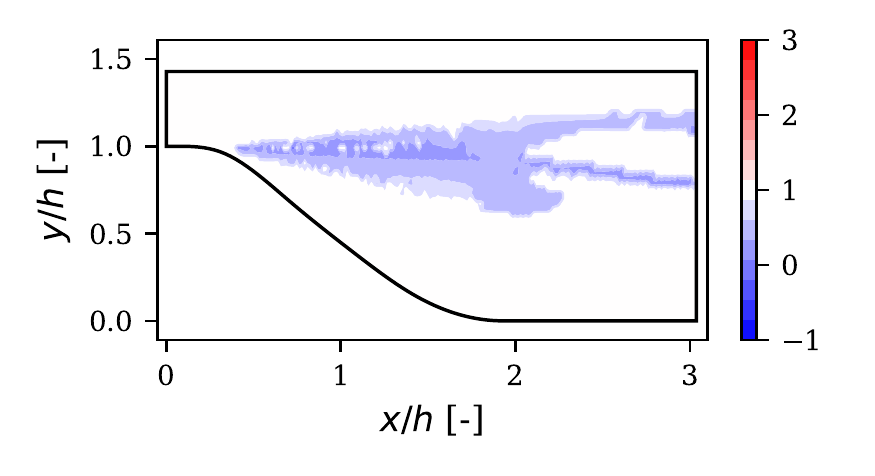}
        \caption{GPEs $\beta_c$ field.}
        \label{fig:ph_beta_gpe}
    \end{subfigure}
    \begin{subfigure}[b]{0.475\textwidth}
        \centering
        \includegraphics[width=\textwidth]{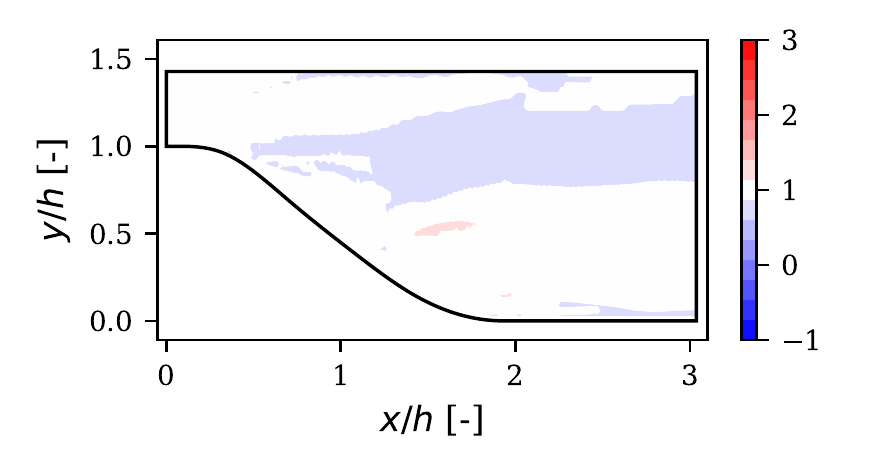}
        \caption{DEs $\beta_c$ field.}
        \label{fig:ph_beta_de}
    \end{subfigure}
    
    \begin{subfigure}[b]{0.475\textwidth}
        \centering
        \includegraphics[width=\textwidth]{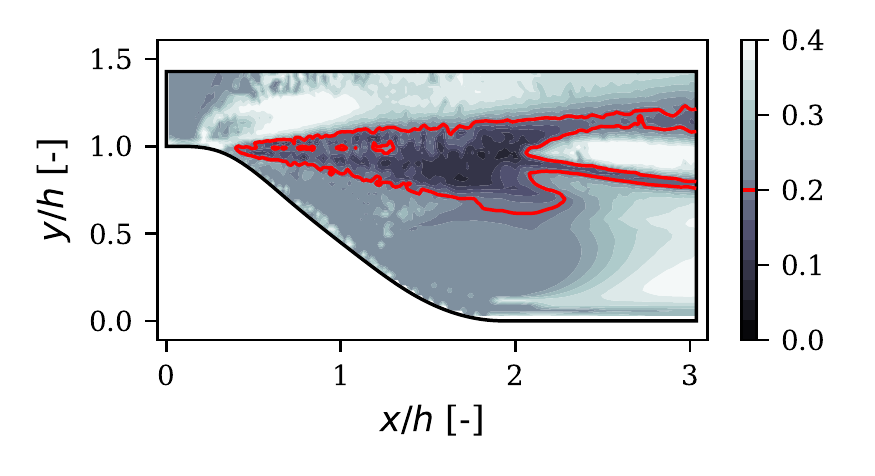}
        \caption{GPEs $\sigma^*$.}
        \label{fig:ph_sigma_gpe}
    \end{subfigure}
    \begin{subfigure}[b]{0.475\textwidth}
        \centering
        \includegraphics[width=\textwidth]{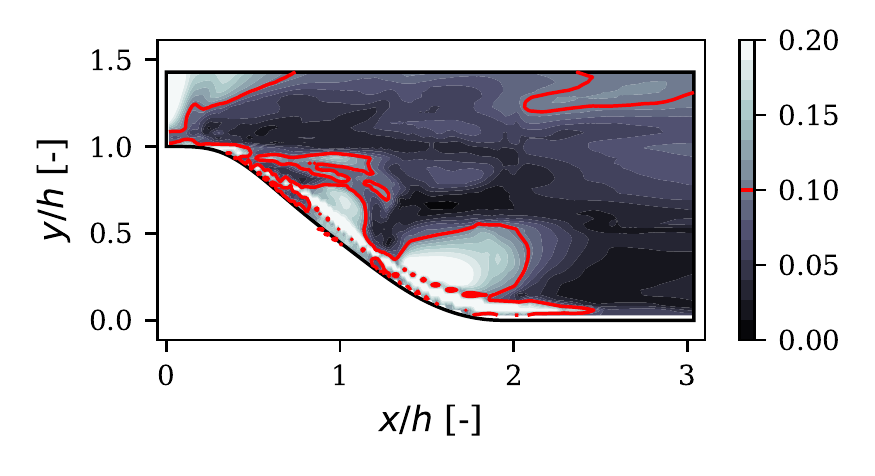}
        \caption{DEs $\sigma^*$.}
        \label{fig:ph_sigma_de}
    \end{subfigure}
    \caption{Periodic hills predictions.}
    \label{fig:ph_beta}
\end{figure}

A region of reduced $\beta_c$ that originates from the separation point of the flow extending downstream around the shear layer can be seen in the inverted results of Figure \ref{fig:ph_beta_inv}. The subsequent increase in $k$ encourages mixing of the shear layer as evident in the velocity profiles which are more spread out in Figure \ref{fig:ph_vel_nut}. The $\nu_t$ profiles show 2 peaks, the first one around $x/h=0.5$ under the crest of the hill, and the second around $x/h=2$ in the region above the hills. The inversion resulted in a more pronounced first peak as seen in Figure \ref{fig:ph_vel_nut}.

The GPEs were active in similar regions of the inverted field with the corrections originating further off the wall. The DEs was active in a larger region with most of the predicted $\beta_c$ values around the value of $1$. The resultant $\beta_c$ field of DEs where the corrections were actually active was similar to GPEs but at a lower magnitude. The increase in the first peak of $\nu_t$ was well predicted by the GPEs with the DEs performing reasonably well. As a result, GPEs predicted a better velocity field compared to DEs, with the velocity profile close to the inverted results. Note that in Figure \ref{fig:ph_sigma_gpe}, the GPE variance is relatively high in the freestream where $\mu^*\approx1$, i.e. the model predicts inactive $\beta_c$. This is to be expected given the preparation of the training data, however, this effect is not noticeable for the predicted uncertainty of the DEs in Figure \ref{fig:ph_sigma_de}. 
\FloatBarrier

\subsection{Zero pressure gradient flat plate}
\begin{figure}[h]
    \centering
    \begin{subfigure}[c]{0.475\textwidth}
        \centering
        \includegraphics[width=\textwidth]{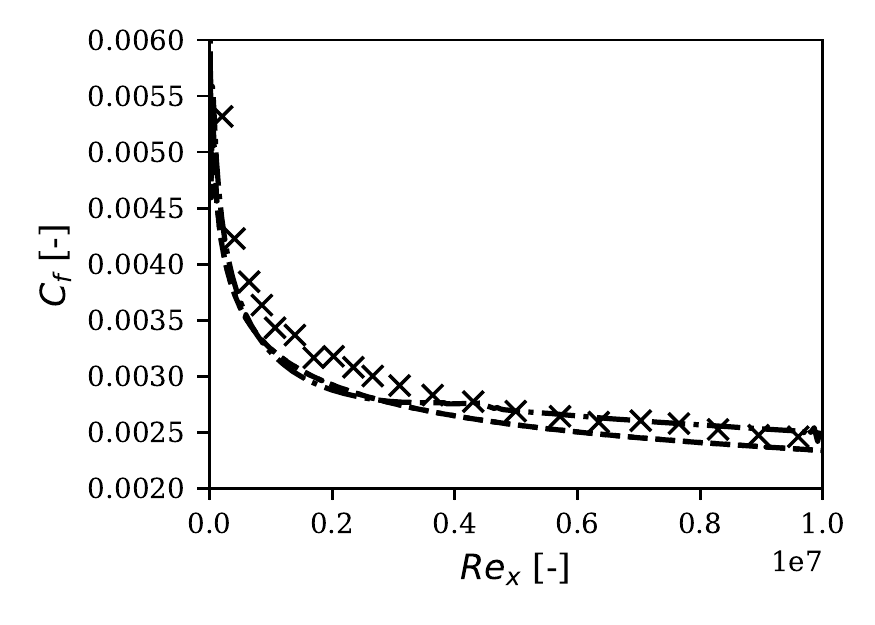}
        \caption{Skin friction.}
        \label{fig:zpg_cf}
    \end{subfigure}
    \begin{subfigure}[c]{0.415\textwidth}
        \centering
        \includegraphics[width=\textwidth]{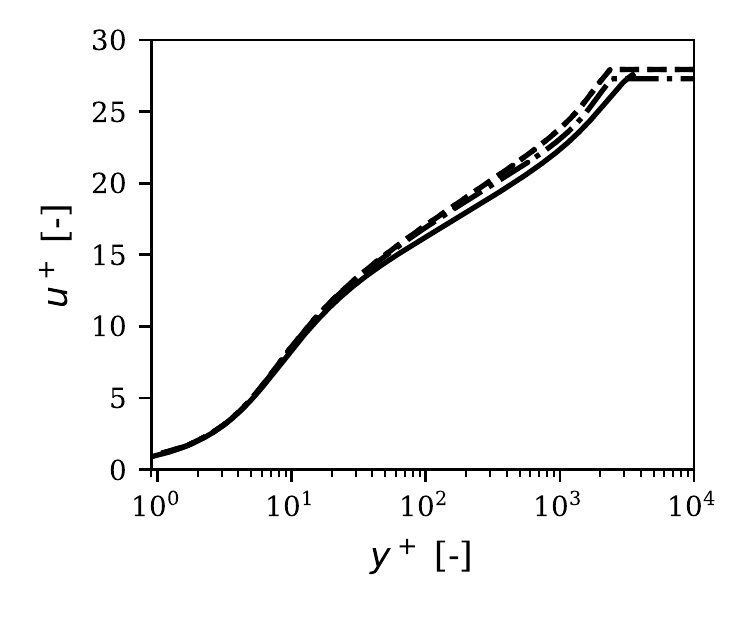}
        \caption{Velocity profile at $Re_x=4.85e6$.}
        \label{fig:zpg_uplus}
    \end{subfigure}
    \caption{Results of zero pressure gradient flat plate. Experiment (\markercross), theory (\full), uncorrected (\dotted), GPEs (\dashed), DEs (\dotdash). Uncorrected and GPEs results overlap.}
    \label{fig:zpg_profiles}
\end{figure}

The simulation domain in this case extends from $x=-0.33m$ to $x=2m$ with the wall starting at $x=0m$. The Reynolds number per unit length is $5\times 10^6$, with the velocity profile measured at $x=0.97m$ corresponding to $Re_{\theta}=1\times 10^5$. A grid of $545\times 770$ was used with a resultant maximum $y^+$ of 0.9 at the first cell. The skin friction was compared against experimental results from that of Wieghardt \cite{wieghardt} in Figure \ref{fig:zpg_cf}. The velocity profile is expected to follow the law of the wall with Coles relation \cite{coles_1956} used as the theoretical reference solution in Figure \ref{fig:zpg_uplus}. Most well designed turbulence models are expected to yield good results on this case and it was raised by Rumsey et al. \cite{gen3} that a NN augmented SA model trained on an adverse pressure gradient flat plate case was predicting spurious $\beta_c$ in the zero pressure case causing undesirable deviations in the skin friction profile. Compared to the training cases, this case does not have any pressure gradients and flow curvature hence the ML models should ideally not be active.

\begin{figure}[h]
    \centering
    \begin{subfigure}[b]{0.475\textwidth}
        \centering
        \includegraphics[width=\textwidth]{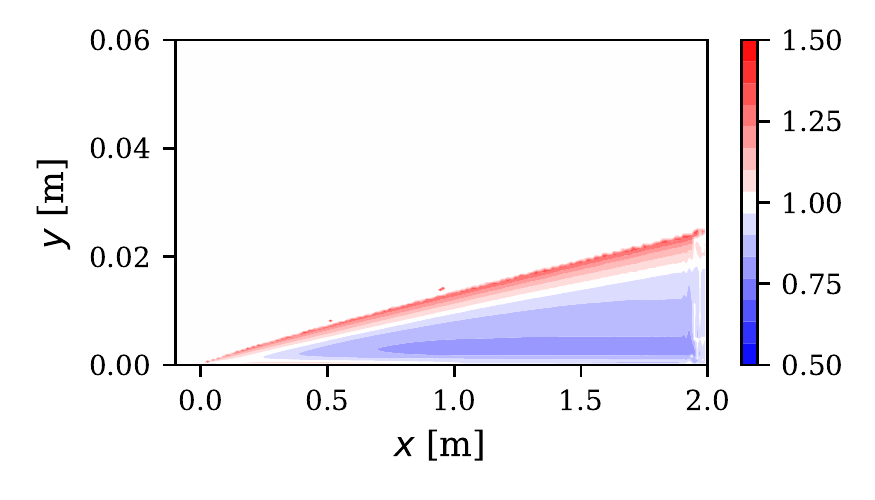}
        \caption{DEs $\beta_c$ field.}
        \label{fig:zpg_de_beta}
    \end{subfigure}
    
    \begin{subfigure}[t]{0.475\textwidth}
        \centering
        \includegraphics[width=\textwidth]{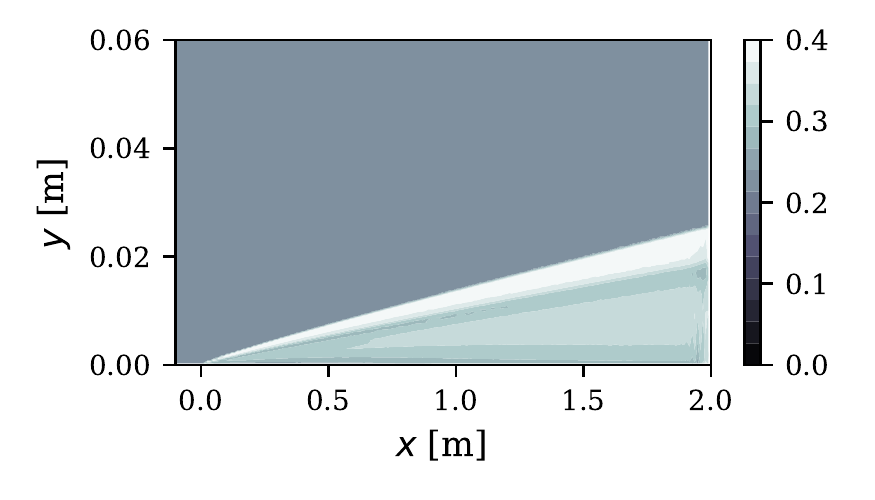}
        \caption{GPEs $\sigma^*$}
        \label{fig:zpg_gpe_sigma}
    \end{subfigure}
    \begin{subfigure}[t]{0.475\textwidth}
        \centering
        \includegraphics[width=\textwidth]{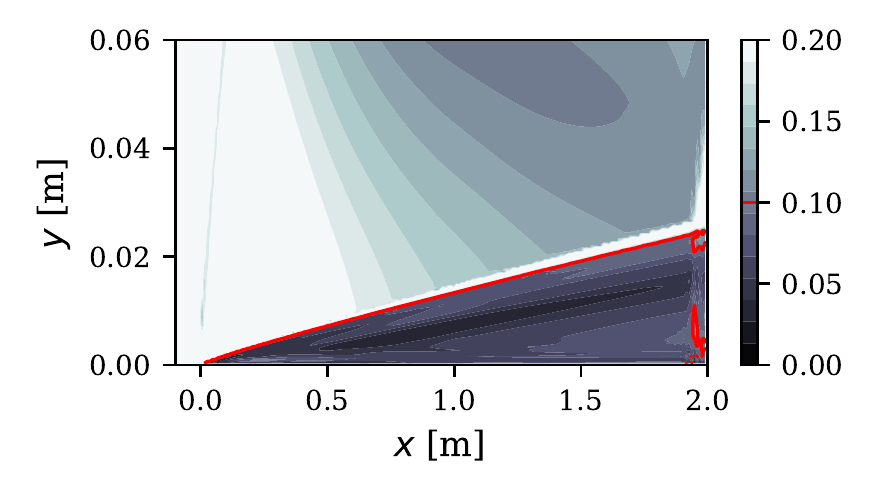}
        \caption{DEs $\sigma^*$}
        \label{fig:zpg_de_sigma}
    \end{subfigure}
    
    \caption{Predicted variables from ML models on zero pressure gradient flat plate case. $\beta_c$ and red contours from GPE not shown as flow field was uncorrected everywhere.}
    \label{fig:zpg}
\end{figure}

The GPEs predicted no activation, with the skin friction and velocity profile overlapping that of the uncorrected model in Figure \ref{fig:zpg_profiles}. Figure \ref{fig:zpg} shows that the DEs predicted $\beta_c$ values in the boundary layers all along the wall with a corresponding increase in $C_f$ from $Re_x=0.3$ onwards. The velocity profile also diverged from the uncorrected model in the log law region as seen in Figure \ref{fig:zpg_uplus}. Although the dimensionless velocity profile of the DEs shifts towards the theoretical relation, the overall behaviour of the DEs is undesirable as it altered the skin friction profile. 

\FloatBarrier

\subsection{NACA 0012 airfoil}
The ML model was applied to the NACA 0012 airfoil in the attached flow regime at $0^{\circ}$, $10^{\circ}$ and $15^{\circ}$ angles of attack. The Reynolds number was $Re_c=6\times 10^6$ with the maximum mesh $y^+=0.3$. Table \ref{table:airfoil_lift_drag} shows the lift and drag values. The uncorrected model was able to predict the lift coefficient well, with all $C_L$ values within $1.5\%$ of the experimental data from Ladson \cite{Ladson1988EffectsOI}. $C_D$ values from the uncorrected model were within $5\%$ of the experiment value, with the error within $0.2\%$ at zero angle of attack. Hence it is desirable for the ML models not to apply any corrections here.

\begin{table}[h]
\begin{center}
    \begin{tabular}{ |c|c|c|c|c|c|c|c|c| } 
    \hline
     & \multicolumn{2}{|c|}{Experiment} & \multicolumn{2}{|c|}{Uncorrected} & \multicolumn{2}{|c|}{GPEs} & \multicolumn{2}{|c|}{DEs} \\ 
    \hline
    $\alpha [^{\circ}]$ & $C_D$ & $C_L$ & $C_D$ & $C_L$ & $C_D$ & $C_L$ & $C_D$ & $C_L$ \\
    \hline
    0 & 0.00810 & 0 & 0.00809 & 0 & 0.00809 & 0 & 0.0084 & 0 \\ 
    10 & 0.01190 & 1.0586 & 0.01129 & 1.0739 & 0.01131 & 1.0774 & 0.0117 & 1.0687 \\
    15 & 0.01830 & 1.4938 & 0.01918 & 1.5099 & 0.01900 & 1.5320 & 0.0199 & 1.5030 \\
    \hline
    \end{tabular}
\caption{Lift and drag coefficients for NACA0012 airfoil compared to experimental data.}
\label{table:airfoil_lift_drag}
\end{center}
\end{table}

\begin{table}[h]
\begin{center}
    \begin{tabular}{ |c|c|c|c|c| } 
    \hline
    & \multicolumn{2}{|c|}{GPEs} & \multicolumn{2}{|c|}{DEs} \\ 
    \hline
    $\alpha [^{\circ}]$ & $\Delta C_D [\%]$ & $\Delta C_L [\%]$ & $\Delta C_D [\%]$ & $\Delta C_L [\%]$ \\
    \hline
    0 & 0.08 & $-$ & 3.34 & $-$ \\
    10 & 0.12 & 0.33 & 3.38 & -0.48 \\ 
    15 & -0.91 & 1.47 & 3.71 & -0.45 \\
    \hline
    \end{tabular}
\caption{Percentage difference in lift and drag coefficient of ML models compared to uncorrected model.}
\label{table:airfoil_delta}
\end{center}
\end{table}

\begin{figure}[h]
    \centering
    \begin{subfigure}[b]{0.475\textwidth}
        \centering
        \includegraphics[width=\textwidth]{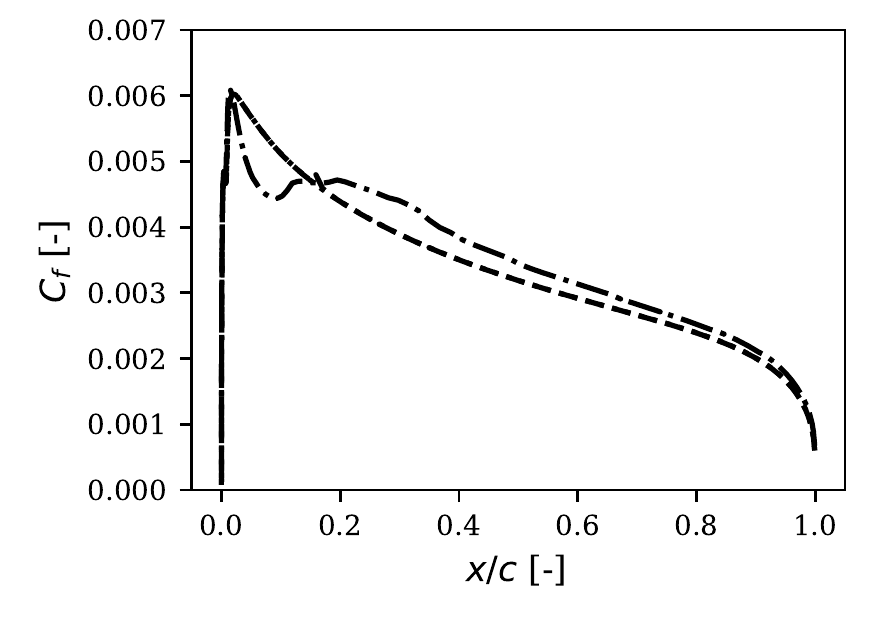}
        \caption{$C_f$ at $0^{\circ}$ angle of attack.}
        \label{fig:airfoil_00deg_cf}
    \end{subfigure}
    \begin{subfigure}[b]{0.475\textwidth}
        \centering
        \includegraphics[width=\textwidth]{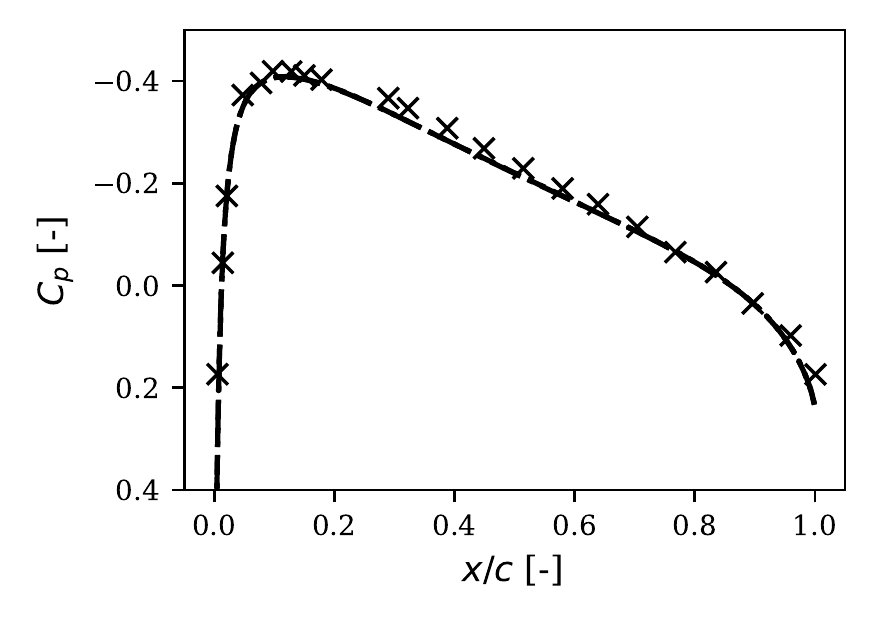}
        \caption{$C_p$ at $0^{\circ}$ angle of attack.}
        \label{fig:airfoil_00deg_cp}
    \end{subfigure}
    
    \begin{subfigure}[b]{0.475\textwidth}
        \centering
        \includegraphics[width=\textwidth]{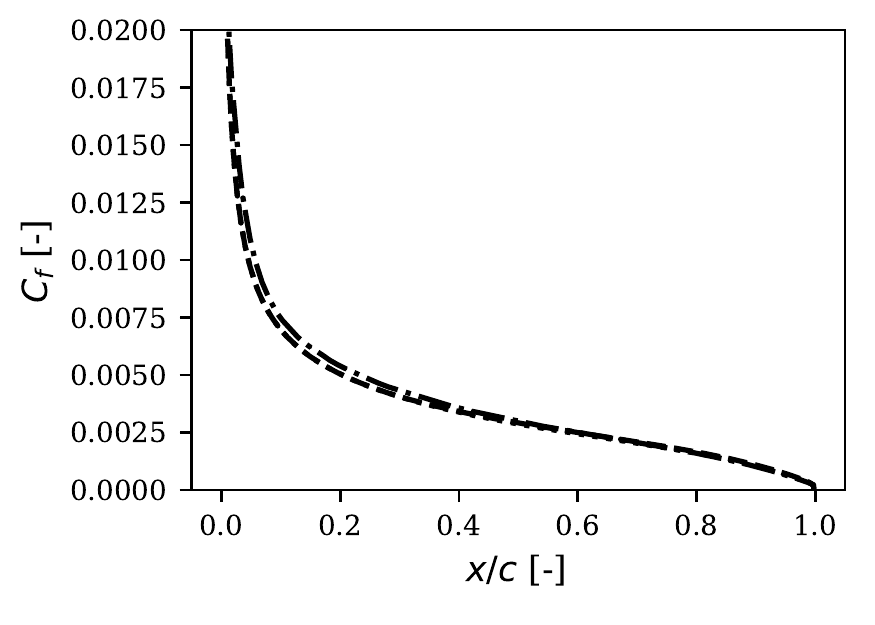}
        \caption{$C_f$ at $10^{\circ}$ angle of attack.}
        \label{fig:airfoil_10deg_cf}
    \end{subfigure}
    \begin{subfigure}[b]{0.475\textwidth}
        \centering
        \includegraphics[width=\textwidth]{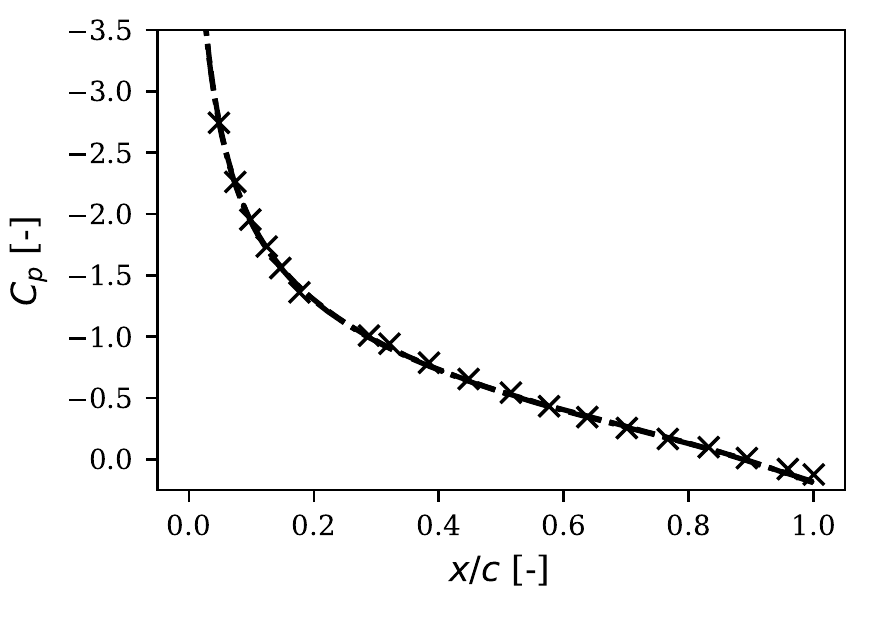}
        \caption{$C_p$ at $10^{\circ}$ angle of attack.}
        \label{fig:airfoil_10deg_cp}
    \end{subfigure}

    \begin{subfigure}[b]{0.475\textwidth}
        \centering
        \includegraphics[width=\textwidth]{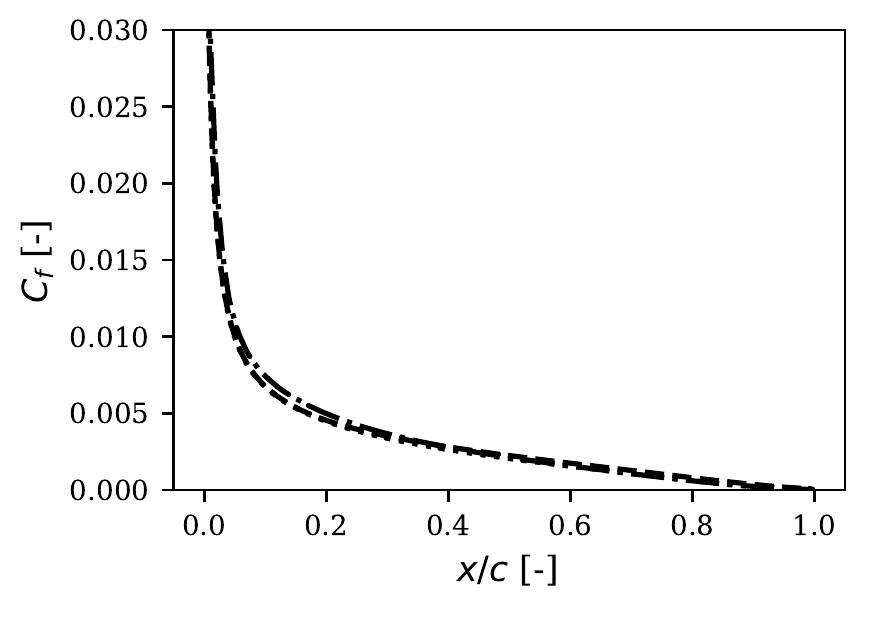}
        \caption{$C_f$ at $15^{\circ}$ angle of attack.}
        \label{fig:airfoil_15deg_cf}
    \end{subfigure}
    \begin{subfigure}[b]{0.475\textwidth}
        \centering
        \includegraphics[width=\textwidth]{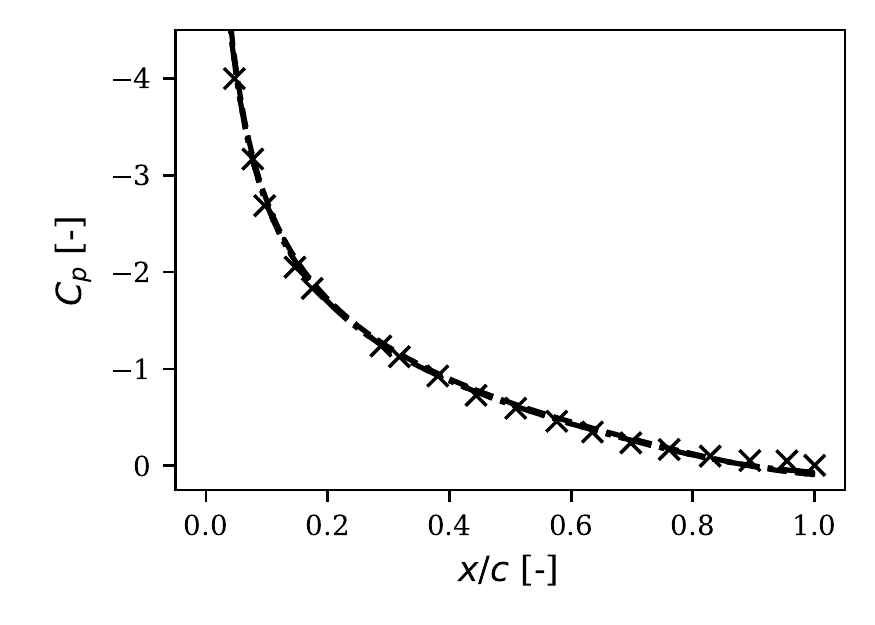}
        \caption{$C_p$ at $15^{\circ}$ angle of attack.}
        \label{fig:airfoil_15deg_cp}
    \end{subfigure}
    
    \caption{Surface variables on suction surface of airfoil at various angles of attack. Experiment (\markercross), uncorrected (\dotted), GPEs (\dashed), DEs (\dotdash). Results from GPEs overlap that of uncorrected results for all plots except $C_p$ at $15^{\circ}$ angle of attack.}
    \label{fig:airfoil_surf_var}
\end{figure}

The GPEs at $0^{\circ}$ and $10^{\circ}$ had lift and drag coefficient values very close to the uncorrected model ($<0.5\%$ as shown in Table \ref{table:airfoil_delta}). At $15^{\circ}$ it predicted a $1.5\%$ higher $C_L$ value. Plotting the $C_p$ profiles at that angle it can be seen in Figure \ref{fig:airfoil_15deg_cp} that they compare well to the experimental data from Gregory and O'Reilly \cite{gregory_1970}. The $1.5\%$ increase in $C_L$ from the GPEs stem from a very slight shift in $C_p$ on the suction surface which is barely visible in Figure \ref{fig:airfoil_15deg_cp}. Figure \ref{fig:airfoil_surf_var} also shows the skin friction on the suction surface. As suggested in Table \ref{table:airfoil_delta}, the GPEs $C_f$ profile overlaps that of the uncorrected model.

The DEs predicted a different $C_f$ profile from the uncorrected model. At zero angle of attack there was a sudden drop in skin friction around the suction peak, followed by an overshoot. At $10^{\circ}$ and $15^{\circ}$, DEs predicted a higher $C_f$ from the leading edge up to $x/c=0.3$. The different $C_f$ profiles leads to a change in drag coefficient of up to around $4\%$ in the DEs compared to the uncorrected model (Table \ref{table:airfoil_delta}).

\FloatBarrier

\section{Conclusions}
% If ML informed turbulence models are to be incorporated within the design process of new aeromechanical components, it is essential that the uncertainty of the model is accounted for and that the prediction reverts to baseline if the ML model is extrapolating to flow physics beyond its training data. This consideration has motivated the development of the probabilistic machine learning method presented here. The proposed ensemble method has the advantage of allowing data from heterogeneous sources to be easily combined. Ensembles of both neural networks and Gaussian Process Emulators were studied, however, as might be expected, the fully probabilistic approach of the GPEs gave a better estimate of the uncertainty than the deep ensembles, where the uncertainty is added \emph{post-hoc}. As a consequence, the GPE ensemble was found to perform better when applied to unseen geometries compared to the Deep Ensembles, reverting to baseline for test cases displaying physics beyond those of the training data.

% While the results presented here are promising, there are several directions in which the method could be extended. For instance, the training dataset could be broadened to include a wider range of flow physics, for example including rotational or internal flows. Increasing the size of the training dataset in this way would also introduce challenges from a Data Science standpoint concerning how best to partition the dataset and structure the ensemble. 

A probabilistic machine learning approach was taken in this work and it has shown to reduce unwanted corrections, leading to more robust predictions. This was achieved through only accepting predictions from the model above a certain confidence. Both GPEs and DEs were able to identify areas of the flow field which needed corrections for cases similar to the training data set (i.e. h42 bump and periodic hills). GPEs were more successful in reverting the turbulence model to baseline for cases which did not include similar flow physics (i.e. zero pressure gradient flat plate and NACA 0012 airfoil).

The success of GPEs is due to the richer probabilistic theory underpinning it. Through an ensemble of GPEs, heterogeneous sources are easily combined with useful results extracted from different components of its variance: $\sigma_{\sigma}^*$ indicates where the model is extrapolating and $\sigma_{\mu}^*$ indicates uncertainty within the training data set.

The modular approach of GPE ensemble allows one to add to the model easily. Training cases including a wider variety of flow phenomena such as rotational and wall bounded flows could be added. However, when the number of training cases increase beyond a certain amount (i.e. tens or hundreds), the question of how best to partition the cases across a limited number of submodels, and how to structure them, becomes important. As shown in Section 4.1, non-unique mappings exists even for cases with similar flow physics. A question remains as to whether a unique enough mapping exists with training sets containing a larger range of flow physics. This behaviour is partly dependent on the feature set chosen, which is also an area worth exploring to improve the mapping.

\section*{Acknowledgement}
We would like to thank Vittorio Michelassi for his helpful comments. We would also like to thank Christopher Rumsey and his group for the discussion and feedback.

\bibliography{probML_CFD.bib}

%\appendix
%\section{Selecting tolerance}

%\begin{figure}[h]
%    \centering
%    \includegraphics[width=0.6\textwidth]{images/GPE/emergent_behav.pdf}
%    \caption{$L_1$ error against ensemble uncertainty for the training dataset}
%    \label{fig:sig_tol}
%\end{figure}
\end{document}